\documentclass{article}
\pdfoutput=1

\usepackage{amsthm}
\usepackage{amsmath,amsfonts}
\usepackage{dsfont}
\usepackage{bbm}
\usepackage{natbib}
\usepackage{graphicx}
\usepackage{enumerate}
\usepackage{url} % not crucial - just used below for the URL 
\usepackage{subcaption} % the subfigure and subfig packages are deprecated
\usepackage{algorithm}
\usepackage{algpseudocode}

\newtheorem{theorem}{Theorem}

  \title{Scalable Bayesian Inference for the Inverse Temperature of a Hidden Potts Model}
  \author{Matthew T. Moores\thanks{National Institute for Applied Statistics Research Australia, School of Mathematics and Applied Statistics, University of Wollongong, NSW 2522, Australia} \and Geoff K. Nicholls\thanks{Department of Statistics, University of Oxford, 24-29 St Giles Oxford OX1 3LG, UK} \and Anthony N. Pettitt\thanks{Australian Research Council Centre for Excellence in Mathematical and Statistical Frontiers (ACEMS)}~\thanks{Mathematical Sciences School,  Queensland University of Technology, Brisbane, Queensland 4001, Australia} \and Kerrie Mengersen\footnotemark[2]~\footnotemark[3]}

\begin{document}
\maketitle

\begin{abstract}
The inverse temperature parameter of the Potts model governs the strength of spatial cohesion and therefore has a major influence over the resulting model fit. A difficulty arises from the dependence of an intractable normalising constant on the value of this parameter and thus there is no closed-form solution for sampling from the posterior distribution directly. There are a variety of computational approaches for sampling from the posterior without evaluating the normalising constant, including the exchange algorithm and approximate Bayesian computation (ABC). A serious drawback of these algorithms is that they do not scale well for models with a large state space, such as images with a million or more pixels. We introduce a parametric surrogate model, which approximates the score function using an integral curve. Our surrogate model incorporates known properties of the likelihood, such as heteroskedasticity and critical temperature. We demonstrate this method using synthetic data as well as remotely-sensed imagery from the Landsat-8 satellite. We achieve up to a hundredfold improvement in the elapsed runtime, compared to the exchange algorithm or ABC. An open source implementation of our algorithm is available in the \textsf{R} package {\bf bayesImageS}.
\end{abstract}

\section{Introduction}

Markov random field (MRF) models have seen widespread use in image analysis since their introduction by \citet{Besag1974}, as surveyed by \citet{Winkler2003} and \citet{Li2009}. A MRF is a generalisation of the Markovian dependence structure to more than one dimension: satellite imagery has two spatial dimensions, while medical images such as computed tomography (CT) are three-dimensional. The hidden \citet{Potts1952} model employs a latent MRF on discrete states to describe spatial dependence between adjacent neighbours. The degree of dependence in the model is governed by a parameter $\beta$, known as the inverse temperature due to its origin in statistical physics \citep{Huang2010}. It is difficult to set this parameter by trial and error, particularly for noisy images. Rather than using a fixed value, it would be preferable to estimate $\beta$ as part of the model. However, the intractable normalising constant of the Potts model depends on the value of the inverse temperature. This means that there is no closed-form solution for sampling from its posterior distribution, since the Metropolis-Hastings (M-H) ratio cannot be computed directly.

\citet{Moeller2006} introduced an auxiliary variable method that gives an exact MCMC algorithm for the special case of a 2 component Potts model, also known as an Ising model. The exchange algorithm of \citet{Murray2006} is a variant of this exact method that avoids the need for a fixed estimate of $\beta$. A drawback of these algorithms is that they require unbiased sampling from the stationary distribution of the Potts model. This is possible for $k=2$ or 3 labels using coupling from the past \citep[CFTP;][]{Propp1996}, perfect slice sampling \citep{Mira2001}, or bounding chains \citep{Huber2003,Huber2016}. The recursive algorithm of \citet{Reeves2004} can also be used to obtain an exact sample \citep{Friel2007}, but only if the lattice is very small. \citet{McGrory2009} reported that the time required for CFTP increased sharply for larger values of $\beta$. The available methods for perfect simulation can be computationally prohibitive for practical applications. For this reason, \citet{Cucala2009} substitute 500 iterations of Gibbs sampling on the auxiliary variable to produce an approximate sample from its stationary distribution. This approximate exchange algorithm (AEA) can also be applied for Potts models with $k > 3$, where the state space $\mathcal{Z}$ cannot be partially ordered. 

Like the exchange algorithm, approximate Bayesian computation (ABC) uses an auxiliary variable to decide whether to accept or reject the proposed value of $\beta$. ABC for the Potts model was introduced by \citet{Grelaud2009}. They observed that these models form a natural exponential family and hence possesses a sufficient statistic, $\mathrm{S}(\mathbf{z})$. \citet{Everitt2012} combined AEA with particle MCMC and also implemented ABC with sequential Monte Carlo (SMC-ABC) for the Ising model. \citeauthor{Everitt2012} found that the computational costs of both AEA and ABC were dominated by simulation of the auxiliary variable. While exact inference is theoretically possible, its applicability is limited to datasets with fewer than a thousand pixels. Comparisons such as \citet{McGrory2009,Moores2014a} and \citet{Moores2014} have shown that auxiliary variable methods such as AEA and ABC are infeasible for the scale of data that is regularly encountered in image analysis.

\begin{table}
\centering
\begin{tabular}{rrrr}
\hline
Number & Satellite & CT slices & HD Video  \\
of pixels &  (900m$^2$/px) & (512$\times$512) & 1080p \\\hline
$4^6$    & $3.7 \textrm{km}^2$ & \dots & \dots \\
$7^6$ & $105.9 \textrm{km}^2$ & $0.4$ & \dots \\
$10^6$ & $900.0 \textrm{km}^2$ & $3.8$ & $0.5$ \\
$15^6$ & $10251.6 \textrm{km}^2$ & $43.5$ & $5.5$ \\ \hline
\end{tabular}
\caption{Scale of common types of images, including the area covered by a satellite image, the number of axial slices of a CT scan, and the number of frames of HD video.} \label{t:scale}
\end{table}
Images containing multiple megapixels are now commonplace, from digital photography and high-definition (HD) video to medical imaging and remote sensing. Table~\ref{t:scale} illustrates how the number of pixels translates to real world scale for various types of images. There is often the need to use multiple images to cover the spatial (and temporal) extent of an imaging study. Multispectral images from the Landsat-7 \citep{NASA2011} and Landsat-8 \citep{USGS2014} satellites have a spatial resolution of 30 metres per pixel (area of 900m$^2$). A Landsat image covers an area of approximately 170km north-south by 183km east-west. Tomographic reconstructions such as CT scans are usually represented as 3D image stacks, with $512 \times 512$ pixels per slice. The pixel resolution and slice width varies depending on the clinical protocol. A single frame of HD video is typically $1920 \times 1080$ pixels with 24 frames per second (fps), although higher frame rates and resolutions (such as ultra high definition, UHD) are available. The volumes of data involved in video processing necessitate specialised methods, which are beyond the scope of this paper. For examples of Bayesian methods for video analysis, see \citet{Simoncelli1999,Minvielle2010}, and the references therein. The remainder of the discussion will focus on static 2D images.

Scalable inference for intractable likelihoods is an active area of research. Some algorithms have been able to improve their runtime while still targeting the exact posterior distribution. For example, both Russian roulette \citep{Lyne2015} and lazy ABC \citep{Prangle2014} used random truncation of the likelihood computation. \citet{Liang2015} used an auxiliary chain to define an importance sampling (IS) distribution for the exchange algorithm. \citet{Sherlock2015} used $k$-nearest neighbours ($k$NN) combined with delayed acceptance \citep{Christen2005} to accelerate pseudo-marginal MCMC. Compared to the exchange algorithm or ABC, these ``exact-approximate'' methods have provided an improvement in computational efficiency of an order of magnitude or less. This is not always sufficient to meet the practical requirements of real world applications.

For Bayesian inference to be feasible on large image datasets, it is necessary to sacrifice some degree of accuracy in order to achieve further improvements in speed.  A variety of approximate methods have recently been proposed, which provide a tradeoff between accuracy and computational cost. One particular class of methods, known as Bayesian indirect likelihood \citep[BIL; ][]{Drovandi2011a,Drovandi2014}, employs a surrogate model to approximate the likelihood function. BIL accelerates computation by interpolating between previous values of the auxiliary variables. It can also take advantage of massively parallel hardware to initialise the surrogate model using a precomputation step. \citet{Moores2014} used linear interpolation to accelerate ABC for the hidden Potts model. 
\citet{Boland2017} derived a theoretical upper bound on the bias introduced by this and similar piecewise approximations.
\citet{Conrad2014} used local polynomial or Gaussian process (GP) models to accelerate MCMC. \citet{Wilkinson2014} and \citet{Meeds2014} used GP emulation for ABC, while \citet{Drovandi2015a} used a GP in a pseudo-marginal algorithm. \citet{Jaervenpaeae2016} used a heteroskedastic GP model and demonstrated how the output of the precomputation step could be used for Bayesian model choice. The IS \citep{Liang2015} and $k$NN \citep{Sherlock2015} approaches could also be viewed as forms of BIL. These nonparametric surrogate models approximate the likelihood under fairly generic assumptions regarding its functional form.

In this paper, we introduce the parametric functional approximate Bayesian (PFAB) algorithm for the Potts model. This algorithm incorporates a surrogate model that approximates the distribution of the sufficient statistic, therefore it is a type of BIL. The advantage over previous BIL algorithms is that we use a parametric integral curve that reflects the known properties of the intractable likelihood function, providing efficiency gains in comparison to nonparametric alternatives. We estimate the parameters of the function using a parallel precomputation step. This leads to a great improvement in runtime, since there is no longer any need to simulate auxiliary variables during model fitting. Using a simulation study, we demonstrate empirically that this approximation error has very little impact on the resulting estimate of the posterior distribution. We also apply our algorithm to images from the Landsat-8 satellite, where the computational cost of AEA or ABC is prohibitive. The superior scalability of PFAB makes it possible to perform fully Bayesian inference for demanding applications such as this.

The remainder of this article is organised as follows. The synthetic data and satellite imagery are described in Section~\ref{s:data}. We define the hidden Potts model in Section~\ref{s:model} and examine the properties of its intractable normalising constant. Our PFAB algorithm for estimating the inverse temperature is described in Section~\ref{s:methods}. We present our experimental results in Section~\ref{s:results} and conclude the article with a discussion.

\section{Example datasets}
\label{s:data}

\subsection{Synthetic data}
\label{s:data_sim}
The inverse temperature cannot be directly observed, so the only way to measure the accuracy of a method is to use synthetic data with known parameter values. We employ the simulation-based calibration (SBC) method of \citet{Cook2006,Talts2018} to evaluate the quality of our PFAB approximation to the posterior for $\beta$. The central concept of this approach is to 
draw a set of values for the model parameters from the prior distribution, then to simulate data from the generative model for each set of parameters, and finally to fit the model and compare the estimates of the posterior distribution to the known values. Simulation-based methods for validating Bayesian posterior coverage have also been proposed by \citet{Monahan1992,Geweke2004}; and in the context of ABC by \citet{Prangle2014a}. \citeauthor{Cook2006} used posterior quantiles for this comparison, which assumes independent samples \citep{Gelman2017}. Instead, we compute the rank statistics suggested by \citeauthor{Talts2018}, which are more applicable to autocorrelated samples obtained from MCMC methods. %The choice of prior distribution for the inverse temperature is discussed in Section~\ref{s:model}.

We use a fixed number of 5 mixture components to generate 7 datasets containing 20 images each. The images within a dataset all have the same dimensions: the smallest images have $2^6 = 8 \times 8$ pixels; the other image sizes are $3^6 = 27 \times 27$; $4^6 = 64 \times 64$; $5^6 = 125 \times 125$; $6^6 = 216 \times 216$; $7^6 = 343 \times 343$; and $10^6 = 1000 \times 1000$. This enables us to study how the performance of the algorithms changes as the size of the images increases. For $n=5^6$ pixels, we also generate datasets with $k=2, 3$ or $4$ hidden states. This gives us 10 simulated datasets in total, containing 200 synthetic images. After discarding the initial, transient phase of the MCMC sampler as burn-in, we thin the output from PFAB to obtain $L = 175$ approximately-independent samples from the posterior (Sect. 5.1 of \citeauthor{Talts2018} explains why this is necessary). 
%According to Thm 1 of \citeauthor{Talts2018}, the rank statistics of the true parameter values will be uniformly-distributed over the integers $[0, L]$.

\subsection{Satellite remote sensing}
\label{s:data_ndvi}
Abundant image data are available from Earth observation satellites. The size of satellite images and the frequency with which they are generated create a requirement for automated methods of image processing. Remotely-sensed satellite imagery has been used for a variety of purposes, such as  estimating land use \citep{Small2001}, water quality \citep{McClain2009} and economic growth \citep{Henderson2011}. In this study, we aim to classify the pixels in a satellite image according to the type and abundance of vegetation that is present. By labelling the pixels, we can quantify the levels of vegetation  in an area and identify contiguous clusters of forest and parkland. These estimates could be used by environmental scientists to estimate global levels of plant biomass. Land use changes such as vegetation clearing could be detected by monitoring changes in pixel classification over time. 

\begin{figure}
        \centering
        \begin{subfigure}[b]{\linewidth}
                \includegraphics[width=\textwidth]{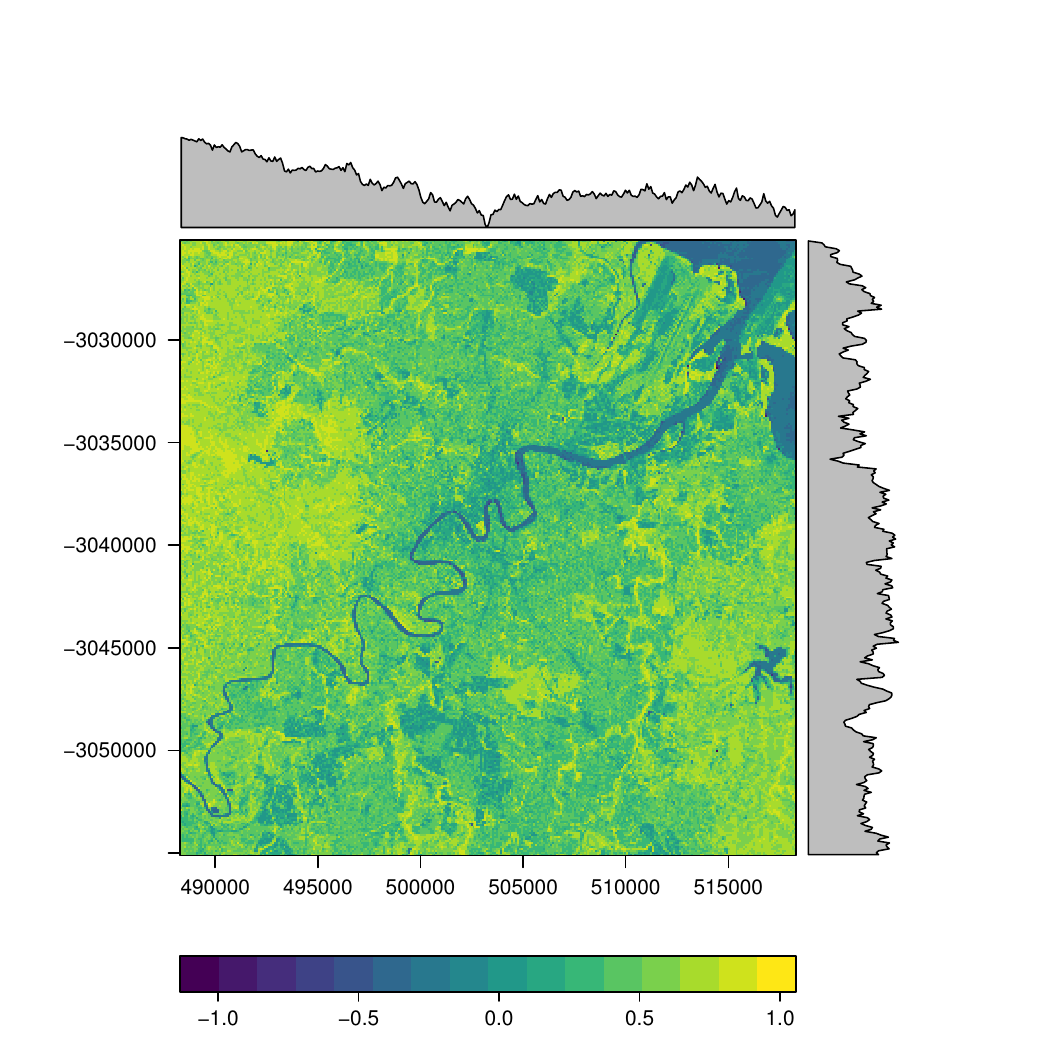}
                \caption{Spatial NDVI distribution.}
                \label{f:ndvi_image}
        \end{subfigure}%
\qquad
        \begin{subfigure}[b]{0.65\linewidth}
                \includegraphics[width=\textwidth]{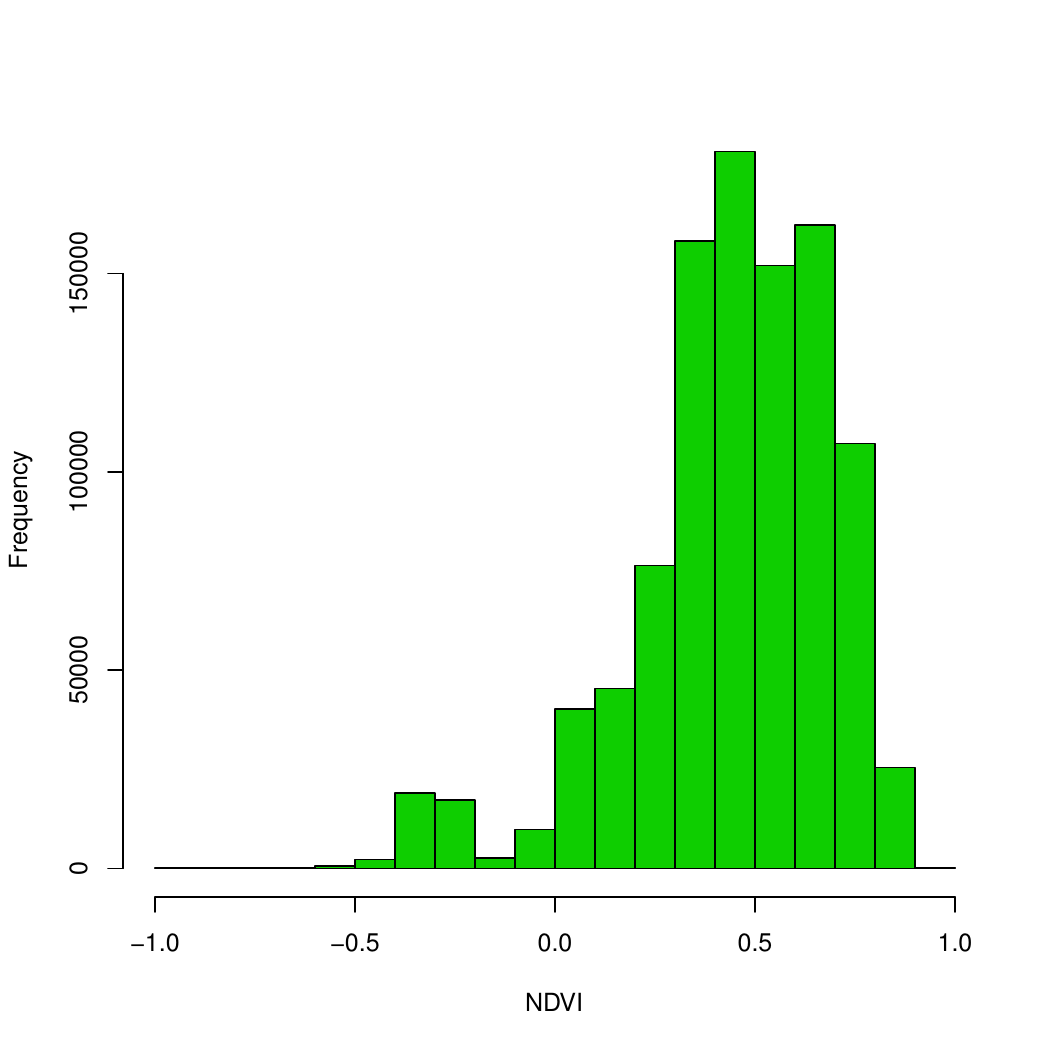}
                \caption{Histogram of NDVI values.}
                \label{f:ndvi_hist}
        \end{subfigure}%
\caption{Normalised difference vegetation index (NDVI) computed from a Landsat-8 satellite image of Brisbane, Australia.}
\label{f:ndvi}
\end{figure}
Metrics such as the normalised difference vegetation index (NDVI) use the ratio of reflectance of red visible light (VIS) and near-infrared (NIR) light as a proxy for the abundance of chlorophyll \citep{Tucker1979,Roy2016}:
\begin{equation}
\label{eq:ndvi}
\mathrm{NDVI} = \frac{NIR - VIS}{NIR + VIS}.
 \end{equation}
Land surface reflectance is estimated from Landsat-8 satellite imagery by correcting for atmospheric effects \citep{Flood2014,Vermote2016}. When converted to NDVI, this produces numbers on a scale from -1 to +1. NDVI is generally used to detect vegetated areas, with an expected value for vegetation being between 0.3 and 0.8. In this study, we aim to classify pixels as vegetation, developed, or water. Using NDVI, we are able to distinguish between urbanised or industrial areas as one sub-category and suburbs as a separate category. We also use separate categories for deep forest and areas of light vegetation. These five categories span the range of NDVI values. However there is no clear separation between them, as illustrated by the histogram in Fig.~\ref{f:ndvi_hist}. A simple thresholding approach would be insufficient to categorise these pixels. Instead, we require a probabilistic approach that takes into account the spatial similarity between pixel labels.

We apply our PFAB algorithm to analyse Landsat-8 imagery of Brisbane, Australia. The Landsat-8 satellite was launched in February 2013 and orbits the Earth every 16 days. To minimise artefacts, we only use imagery with less than 10\% cloud cover. There are 18 such images available for 2015 and 2016. These images were cropped to a region of interest 30 km square, containing one million pixels. This roughly corresponds to the greater metropolitan area. An example of one of these images is shown in Fig.~\ref{f:ndvi_image}. Images from Landsat-8 have 16 bits per pixel, with 55,000 possible values. This is an improvement over Landsat-5 and 7, which produce 8 bit images (256 values). The Landsat-8 bands are also narrower, which provides higher spectral resolution. The VIS band contains wavelengths between 636 and 673 nm, while the NIR band is from 851 to 879 nm \citep{USGS2014}.

\section{Hidden Potts model}
\label{s:model}
Image segmentation can be viewed as the task of labelling the observed pixels $y_i \in \mathbb{R}$, $i = 1,\dots,n$ according to a finite set of discrete states $\mathbf{z} \in \{ 1, \dots, k \}^n$. The hidden Potts model allows for spatial correlation between neighbouring labels in the form of a MRF. The latent labels follow a Gibbs distribution, which is specified in terms of its conditional probabilities:
  \begin{equation}
  \label{eq:Potts}
  p(z_i | z_{\setminus i}, \beta) = \frac{\exp\left\{\beta\sum_{\ell \in \partial(i)}\delta(z_i,z_\ell)\right\}}{\sum_{j=1}^k \exp\left\{\beta\sum_{\ell \in \partial(i)}\delta(j,z_\ell)\right\}},
  \end{equation}
  where $\beta$ is the inverse temperature, $z_{\setminus i}$ represents all of the labels except $z_i$, $\partial(i)$ are the neighbouring pixels of $i$, and $\delta(u,v)$ is the Kronecker delta function. Thus, $\sum_{\ell \in \partial(i)}\delta(z_i,z_\ell)$ is a count of the neighbours that share the same label.

If the pixels in a rectangular 2D lattice with $r$ rows and $c$ columns are indexed row-wise, the nearest (first-order) neighbours $\partial(i)$ are $\ell \in \{ {i-1}, {i-c}, {i+c}, {i+1} \}$, except on the boundary. Pixels situated at the boundary of the image domain have fewer than four neighbours. These neighbourhood relationships are reciprocal, so $h \in \partial(i)$ implies $i \in \partial(h)$. If $\mathcal{E}$ is the set of all unique neighbour pairs or edges $h \sim \ell$ in the image lattice, then $\#\mathcal{E} = 2(n - \sqrt{n})$ for a square lattice or $2 r \cdot c - r - c$ for a rectangular lattice.

The observation equation links the latent labels to the corresponding pixel values:
  \begin{equation}
  \label{eq:obs}
  p(\mathbf{y} | \mathbf{z}, \boldsymbol\theta) = \prod_{i=1}^n p(y_i | z_i, \theta_{z_i}),
  \end{equation}
  where $\theta_{z_i}$ are the parameters that govern the distribution of the pixel values with label $z_i$. The hidden Potts model can thus be viewed as a spatially-correlated generalisation of the finite mixture model \citep{Ryden1998}. \citet{Green2002} used a Poisson likelihood for (\ref{eq:obs}), with intensity $\lambda_{z_i}$. Instead we follow \citet{Geman1984,Alston2007}, and many others in assuming that the pixels with label $j$ share a common mean $\mu_j$ corrupted by additive Gaussian noise with variance $\sigma_j^2$:
  \begin{equation}
  \label{eq:obs2}
y_i | z_i=j, \mu_j, \sigma^2_j \;\sim\; \mathcal{N}\left( \mu_j, \sigma^2_j \right).
  \end{equation}
 
The joint distribution of all of the pixel labels can be expressed in the form of an exponential family, as noted by \citet{Grelaud2009}:
\begin{equation}\label{eq:potts_joint}
p(\mathbf{z} \mid \beta) = \exp\{ \beta \mathrm{S}(\mathbf{z}) - \log \mathcal{C}(\beta) \}.
\end{equation}
The augmented likelihood $p(\mathbf{y},\mathbf{z} | \boldsymbol\theta, \beta)$ can therefore be factorised into $p(\mathbf{y} | \mathbf{z}, \boldsymbol\theta) p(\mathbf{z} \mid \beta)$, where the second factor does not depend on the observed data, but only on the sufficient statistic:
  \begin{equation}
  \label{eq:potts_stat}
\mathrm{S}(\mathbf{z}) = \sum_{i \sim \ell \in \mathcal{E}} \delta(z_i,z_\ell).
  \end{equation}
This statistic represents the total number of like neighbour pairs in the image. The joint posterior is then:
\begin{equation}
\label{eq:joint_post}
p(\boldsymbol\theta, \beta, \mathbf{z} \mid \mathbf{y}) \propto p(\mathbf{y} | \mathbf{z}, \boldsymbol\theta) \pi(\boldsymbol\theta) p(\mathbf{z} | \beta) \pi(\beta),
\end{equation}
where $\pi(\boldsymbol\theta)$ is the joint prior for the parameters of the observation equation \eqref{eq:obs} and $\pi(\beta)$ is the prior for the inverse temperature. We use informative, conjugate priors for $\pi(\mu_j) \sim \mathcal{N}(m_j, s^2_j)$ and $\pi(\sigma^2_j) \sim \mathcal{I}nv\mathcal{G}a(n_j/2, n_j v_j/2)$, based on the properties of the NDVI values:
\begin{eqnarray}
\pi(\boldsymbol\mu) &\sim& \mathcal{N}( \{ -1.0, -0.5, 0.0, +0.5, +1.0 \}, 0.1^2) \label{eq:pr_mu},\\
\pi(\sigma_j^2) &\sim& \mathcal{I}nv\mathcal{G}a(2.5, 0.06) \label{eq:pr_sd},
\end{eqnarray}
The $k=5$ mixture components correspond to different types of land use. In increasing order of their means, these can be classified as: water; urban/industrial; suburban; light vegetation; and dense vegetation. Although these priors are informative for the parameters, there is no direct mapping between the NDVI values and their categorisation. There is too much overlap between the distributions, as illustrated by Fig.~\ref{f:ndvi_hist}. This is why we require the probabilistic labelling approach of the hidden Potts model, rather than a simpler method such as thresholding. For the smoothing parameter $\pi(\beta)$, we follow previous authors including \citet{Moeller2006,McGrory2012,Everitt2012}; and \citet{Lyne2015} in using a uniform prior over a bounded, finite domain. This prior is useful for comparing different algorithms, since it places equal weight on a range of possible values for $\beta$, including both ordered ($\beta > \beta_{crit}$) and disordered ($\beta < \beta_{crit}$) states, as well as the critical point.

The conditional posterior distributions $p(\boldsymbol\theta | \mathbf{z}, \mathbf{y})$ and $p(z_i | z_{\setminus i}, \beta, y_i, \boldsymbol\theta_{z_i})$ can be simulated using Gibbs sampling, but $p(\beta | \mathbf{y}, \mathbf{z}, \boldsymbol\theta)$ involves an intractable normalising constant $\mathcal{C}(\beta)$:
  \begin{eqnarray}
  \label{eq:beta_post}
  p(\beta \mid \mathbf{y}, \mathbf{z}, \boldsymbol\theta) &=&  \frac{p(\mathbf{z} | \beta) \pi(\beta)}{\int_\beta p(\mathbf{z} | \beta) \pi(d \beta)}
\\   \label{eq:beta}
 &\propto& \frac{\exp\left\{ \beta\, \mathrm{S}(\mathbf{z}) \right\}}{\mathcal{C}(\beta)} \pi(\beta).
  \end{eqnarray}
The normalising constant is also known as a partition function in statistical physics. It has computational complexity of $\mathcal{O}(n k^n)$, since it involves a sum over all possible configurations of the labels $\mathbf{z} \in \mathcal{Z}$:
  \begin{equation}
  \label{eq:norm}
\mathcal{C}(\beta) = \sum_{\mathbf{z} \in \mathcal{Z}} \exp\left\{\beta\, \mathrm{S}(\mathbf{z})\right\}.
  \end{equation}
It is infeasible to calculate this value exactly for nontrivial images, thus computational approximations are required.
\begin{figure}
\centering
        \begin{subfigure}{0.45\textwidth}
                \includegraphics[width=\textwidth]{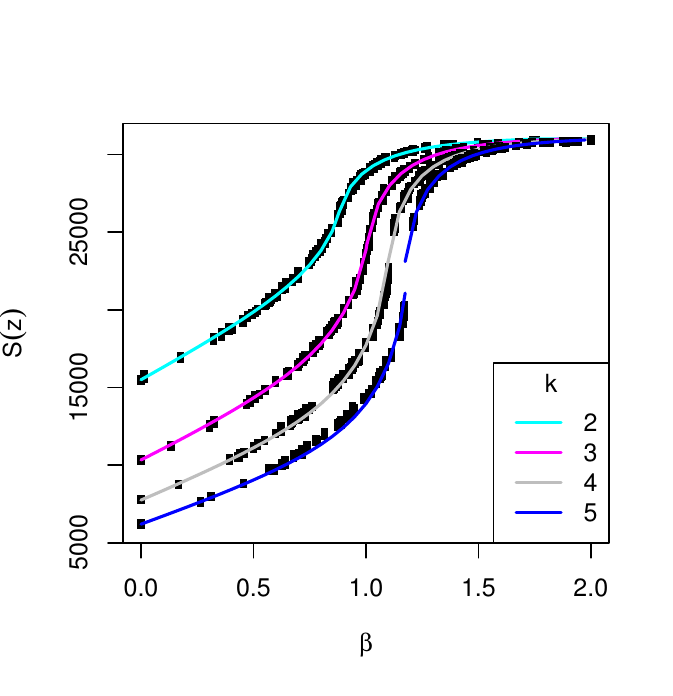}
                \caption{Expectation for $n=5^6$ and increasing values of $k \in \{2, 3, 4, 5\}$.}
                \label{f:exact_exp_k}
        \end{subfigure}
\qquad
        \begin{subfigure}{0.45\textwidth}
                \includegraphics[width=\textwidth]{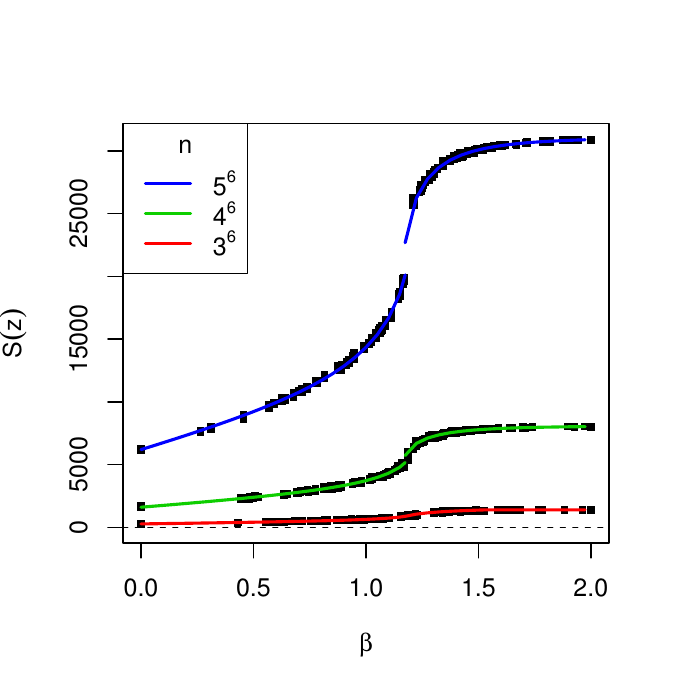}
                \caption{Expectation for $k=5$ and increasing values of $n \in \{3^6, 4^6, 5^6\}$.}
                \label{f:exact_exp_n}
        \end{subfigure}%
\qquad
        \begin{subfigure}{0.45\textwidth}
                \includegraphics[width=\textwidth]{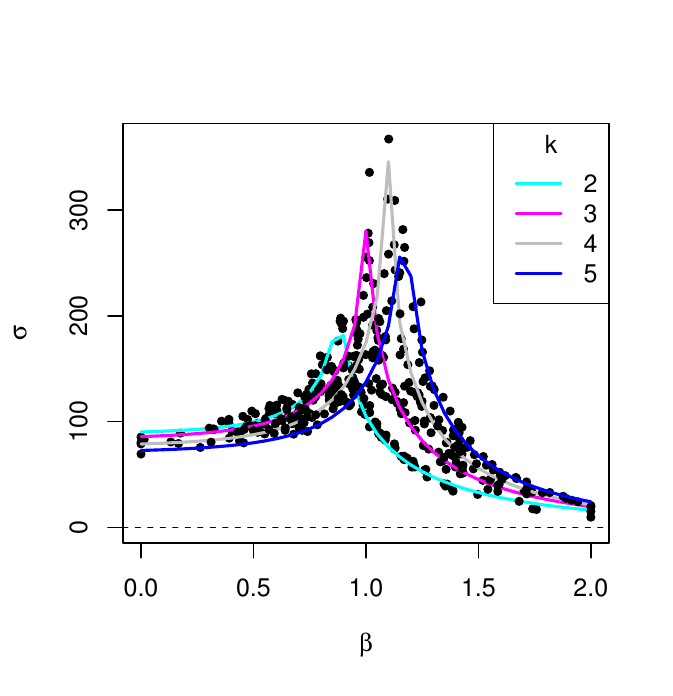}
                \caption{Standard deviation for $n=5^6$ and increasing values of $k \in \{2, 3, 4, 5\}$.}
                \label{f:exact_var_k}
        \end{subfigure}%
\qquad
        \begin{subfigure}{0.45\textwidth}
                \includegraphics[width=\textwidth]{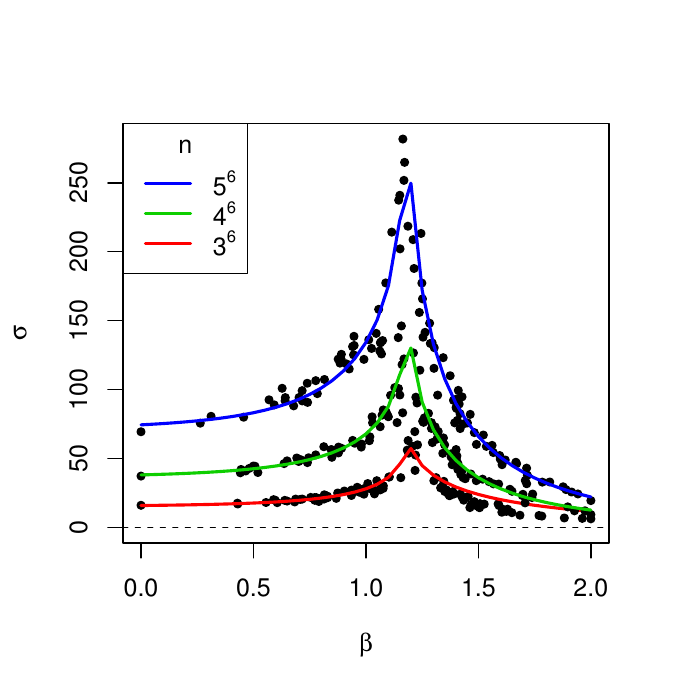}
                \caption{Standard deviation for $k=5$ and increasing values of $n \in \{3^6, 4^6, 5^6\}$..}
                \label{f:exact_var_n}
        \end{subfigure}%
\caption[Distribution of the sufficient statistic of the Potts model]{Approximate mean and standard deviation of $\mathrm{S}(\mathbf{z})$ using Swendsen-Wang for increasing values of the inverse temperature $\beta$, the number of pixels $n$, and the number of unique labels $k$. The black dots are empirical estimates computed from random samples, while the fitted curves illustrate the parametric functional approximation introduced in Section~\ref{s:methods}.}
\label{f:exact_beta}
\end{figure}

The conditional expectation of $\mathrm{S}(\mathbf{z})$ given $\beta$ can be expressed in terms of the normalising constant:
  \begin{equation}
  \label{eq:expSz}
\mathbb{E}_{\mathbf{z} | \beta}[\mathrm{S}(\mathbf{z})] = \frac{\mathrm d}{\mathrm d \beta} \log\{ \mathcal{C}(\beta) \}.
  \end{equation}
 As $\beta$ approaches infinity, all of the pixels in the image are almost surely assigned the same label, thus the expectation of $\mathrm{S}(\mathbf{z})$ approaches the total number of edges $\#\mathcal{E}$ asymptotically, while the variance approaches zero. When $\beta = 0$, (\ref{eq:potts_joint}) simplifies to $\left(\sum_{\mathbf{z} \in \mathcal{Z}} \exp\{0\}\right)^{-1} = k^{-n}$, hence the labels $z_i$ are independent and uniformly-distributed.
 
\begin{theorem} \label{thm:Sz}
The sum over configuration space of the sufficient statistic of the $k$-state Potts model with first-order neighbours is $\sum_{\mathbf{z} \in \mathcal{Z}} \sum_{i \sim \ell \in \mathcal{E}} \delta(z_i,z_\ell) = k^{n-1} \#\mathcal{E}$.
\end{theorem}

\begin{proof}
\begin{eqnarray*}
\sum_{\mathbf{z} \in \mathcal{Z}} S(\mathbf{z}) = \sum_{\mathbf{z} \in \mathcal{Z}} \sum_{i \sim \ell \in \mathcal{E}} \delta(z_i,z_\ell) = \sum_{i \sim \ell \in \mathcal{E}} \sum_{\mathbf{z} \in \mathcal{Z}} \delta(z_i,z_\ell) = \sum_{i \sim \ell \in \mathcal{E}} \frac{k^n}{k} = k^{n-1} \#\mathcal{E}
\end{eqnarray*}
\end{proof}
The size of the configuration space $| \mathcal{Z} |$ is $k^n$. When we reverse the order of summation, we consider only a single pair of indices $i, \ell$ at a time. The Kronecker delta $\delta(z_i,z_\ell) = 1$ for $\frac{1}{k}$ out of the possible values of $z_i$ and $z_\ell$.

\begin{theorem} \label{thm:expSz}
The expectation of the $k$-state Potts model on a rectangular 2D lattice is $\mathbb{E}_{\mathbf{z} | \beta = 0}[\mathrm{S}(\mathbf{z})] = \#\mathcal{E}/k$ when the inverse temperature $\beta = 0$.
\end{theorem}

\begin{proof}
The proof follows from Theorem~\ref{thm:Sz} by noting that $p(\mathbf{z} | \beta = 0) = k^{-n}$ and hence:
\begin{eqnarray*}
\mathbb{E}_{\mathbf{z} | \beta = 0}[\mathrm{S}(\mathbf{z})] &=& \sum_{\mathbf{z} \in \mathcal{Z}} S(\mathbf{z}) \;p(\mathbf{z} \mid \beta = 0) \\
&=& k^{n-1} \#\mathcal{E} \;k^{-n} = \frac{\#\mathcal{E}}{k}.
\end{eqnarray*}
\end{proof}
 
 The score function of the Potts model is the difference between the observed sufficient statistic and its expectation:
 \begin{equation}
 \label{eq:score}
 \frac{\mathrm{d}}{\mathrm{d}\beta} \log\{ p(\mathrm{S}(\mathbf{z}) \mid \beta) \} = \mathrm{S}(\mathbf{z}) - \mathbb{E}_{\mathbf{z} | \beta}[\mathrm{S}(\mathbf{z})].
 \end{equation}
 The variance of the score function, also known as the Fisher information, is given by the expectation of the derivative of \eqref{eq:expSz}:
 \begin{eqnarray} \label{eq:information}
 \mathcal{I}(\beta) &=& \mathbb{E}_{\mathbf{z} | \beta}\left[ \left( \frac{\mathrm{d}}{\mathrm{d}\beta} \log\{ p(\mathrm{S}(\mathbf{z}) \mid \beta) \} \right)^2  \right]\\
 &=& \frac{\mathrm{d^2}}{\mathrm{d}\beta^2} \log\{ \mathcal{C}(\beta) \}. 
 \end{eqnarray}

\begin{theorem} \label{thm:SzSQ}
The sum over configuration space of the square of the sufficient statistic of the $k$-state Potts model on a rectangular 2D lattice is
\begin{equation*}
\sum_{\mathbf{z} \in \mathcal{Z}} \mathrm{S}(\mathbf{z})^2  = k^n \#\mathcal{E} \left(k^{-2}\#\mathcal{E} + k^{-1}(1 - k^{-1})\right).
\end{equation*}
\end{theorem}

\begin{proof}
\begin{eqnarray*}
\sum_{\mathbf{z} \in \mathcal{Z}} \mathrm{S}(\mathbf{z})^2  &=& \sum_{\mathbf{z} \in \mathcal{Z}} \sum_{i \sim \ell \in \mathcal{E}} \sum_{j \sim h \in \mathcal{E}} \delta(z_i,z_\ell) \delta(z_j,z_h)\\
&=&  \sum_{i \sim \ell \ne j \sim h} \sum_{\mathbf{z} \in \mathcal{Z}} \delta(z_i,z_\ell) \delta(z_j,z_h) + \sum_{i \sim \ell = j \sim h} \sum_{\mathbf{z} \in \mathcal{Z}} \delta(z_i,z_\ell)\\
&=& \frac{k^n}{k^2} \#\mathcal{E}^2 + \left(1 - \frac{1}{k}\right)\frac{k^n}{k}\#\mathcal{E}\\
&=& k^n \#\mathcal{E} \left(k^{-2}\#\mathcal{E} + k^{-1}(1 - k^{-1})\right).
\end{eqnarray*}
\end{proof}

\begin{theorem} \label{thm:varSz}
The variance of the $k$-state Potts model on a rectangular 2D lattice is $\mathbb{V}_{\mathbf{z} | \beta = 0}[\mathrm{S}(\mathbf{z})] = \mathcal{I}(0) = \#\mathcal{E} k^{-1} (1 - k^{-1})$ when the inverse temperature $\beta = 0$.
\end{theorem}

\begin{proof}
The proof follows from Theorems~\ref{thm:expSz} and \ref{thm:SzSQ}:
\begin{eqnarray*}
\mathbb{V}_{\mathbf{z} | \beta = 0}[\mathrm{S}(\mathbf{z})] &=& \mathbb{E}_{\mathbf{z} | \beta = 0}[\mathrm{S}(\mathbf{z})^2] - \mathbb{E}_{\mathbf{z} | \beta = 0}[\mathrm{S}(\mathbf{z})]^2\\
&=& \sum_{\mathbf{z} \in \mathcal{Z}} S(\mathbf{z})^2 \;p(\mathbf{z} \mid \beta = 0) - \left(\frac{\#\mathcal{E}}{k}\right)^2\\
&=& k^n \#\mathcal{E} \left(k^{-2}\#\mathcal{E} + k^{-1}(1 - k^{-1})\right) k^{-n} - \frac{\#\mathcal{E}^2}{k^2}\\
&=& \frac{\#\mathcal{E}^2}{k^2} + \#\mathcal{E} k^{-1} (1 - k^{-1}) - \frac{\#\mathcal{E}^2}{k^2} = \#\mathcal{E} k^{-1} (1 - k^{-1}).
\end{eqnarray*}
\end{proof}

\begin{figure}
        \centering
        \begin{subfigure}{0.75\textwidth}
                \includegraphics[width=\textwidth]{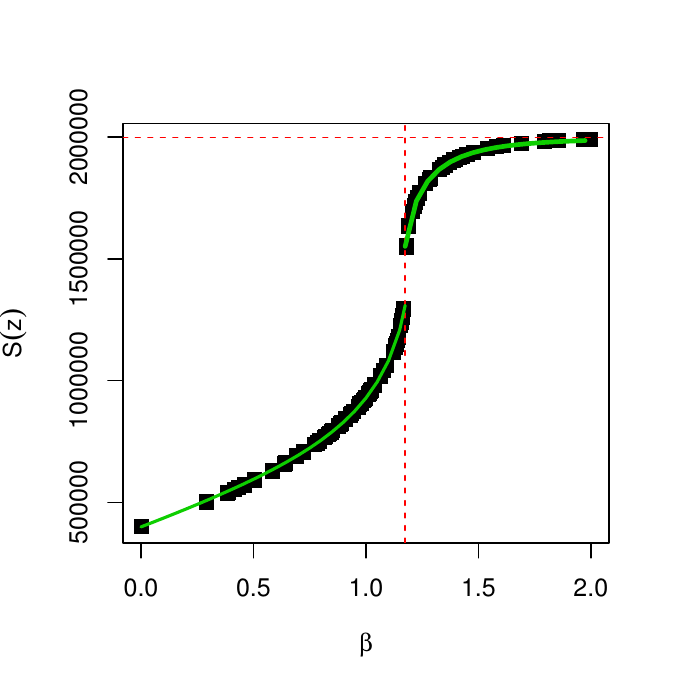}
                \caption{Expectation.}
                \label{f:bcrit2d}
        \end{subfigure}%
\qquad
        \begin{subfigure}{0.75\textwidth}
                \includegraphics[width=\textwidth]{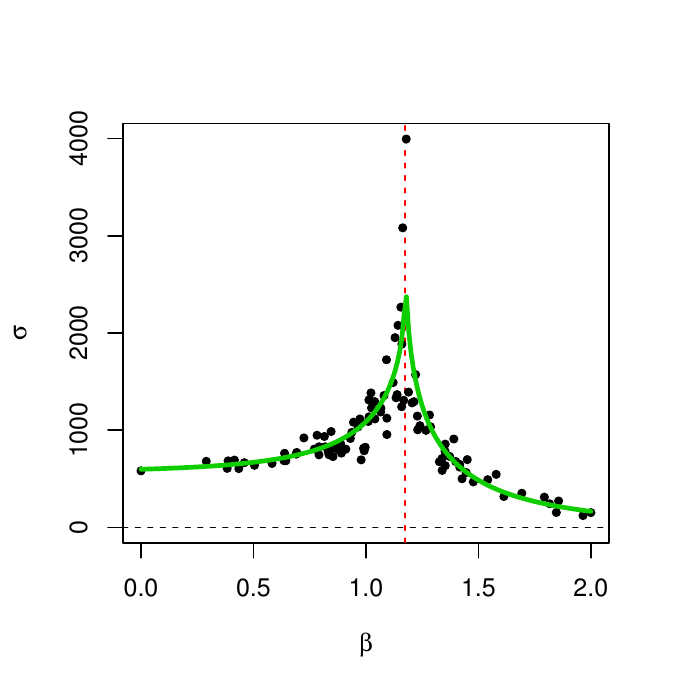}
                \caption{Standard deviation.}
                \label{f:bcrit2d_sd}
        \end{subfigure}%
\caption{Approximate mean and standard deviation of $\mathrm{S}(\mathbf{z})$ using Swendsen-Wang for a 2D image with $k=5$ and $\#\mathcal{E} = 1,998,000$. The vertical, dashed line is the critical value of $\beta$. The horizontal, dashed line is $\#\mathcal{E}$. The black dots are empirical estimates computed from random samples, while the fitted curves illustrate the parametric functional approximation of Section~\ref{s:methods}.}
\label{f:bcrit}
\end{figure}
The Potts model undergoes a phase transition at the critical value of $\beta$, switching from a disordered to an ordered state. \citet{Potts1952} showed that the critical value for a regular 2D lattice can be calculated exactly according to:
  \begin{equation}
  \label{eq:bcrit2d}
\beta_{crit} = \log\left\{1 + \sqrt{k}\right\}.
  \end{equation}
This is the location of the phase transition for a lattice with $r$ rows and infinite columns, so for example the critical value for the images in Figure~\ref{f:exact_beta} is different to (\ref{eq:bcrit2d}). However, the error introduced by a finite boundary diminishes as $n$ increases, due to the finite-dimensional scaling property of the Potts model. Figure~\ref{f:bcrit} shows that (\ref{eq:bcrit2d}) is very accurate in predicting the behaviour of $\mathrm{S}(\mathbf{z})$ for a 2D image with a maximum value of $\mathrm{S}(\mathbf{z})$ for first-order neighbours of 1,998,000. This corresponds to the satellite image described in Section~\ref{s:data_ndvi}, with k=5 mixture components and $\beta_{crit} \approx 1.174$. $\mathrm{S}(\mathbf{z})$ is approximated by simulation using the algorithm of \citet{Swendsen1987}.

Figure~\ref{f:exact_exp_k} illustrates how the location of the phase transition changes for different $k$. \citet{Baxter1973} established that the score function of the Potts model is continuous and smoothly-varying for $k \le 4$. This is known as a second-order phase transition, since there is a sharp change in the variance as shown in Figure~\ref{f:exact_var_k}.  This heteroskedasticity has important implications for the methods discussed in Section~\ref{s:methods}. For $k > 4$ there is a first-order phase transition at the critical point, shown by the discontinuity in Figures \ref{f:exact_exp_k}, \ref{f:exact_exp_n}, and \ref{f:bcrit2d} for $k=5$.  The difference between $\lim_{\beta \nearrow \beta_{crit}}\,\mathbb{E}_{\mathbf{z} | \beta}[\mathrm{S}(\mathbf{z})]$ and  $\lim_{\beta \searrow \beta_{crit}}\,\mathbb{E}_{\mathbf{z} | \beta}[\mathrm{S}(\mathbf{z})]$ is known as the latent heat. This phase transition only emerges as $n$ becomes large, so for example there is no discontinuity for $n=2^6$ even when $k=5$.

\section{Bayesian indirect likelihood}
\label{s:methods}
In AEA and ABC, the conditional distribution $f(\mathrm{S}(\mathbf{w}) \mid \beta)$ of the auxiliary variables or pseudo-data $\mathbf{w}$ is independent of the observed data $\mathbf{y}$ and the labels $\mathbf{z}$. BIL involves constructing a suitable surrogate model $f_A(\mathrm{S}(\mathbf{w}) \mid \phi(\beta))$ to approximate this distribution, using a function $\phi(\beta)$ that we estimate by simulating pseudo-data at a fixed set of values $\{\beta_s\}_{s=1}^S$. This function can be reused across multiple datasets, amortising its computational cost. By replacing $\mathrm{S}(\mathbf{w})$ with our surrogate model, we avoid the need to simulate auxiliary variables during model fitting. In \citet{Moores2014}, it was shown that this approximation could lead to two orders of magnitude improvement in the elapsed runtime for fitting the hidden Potts model, while also improving the convergence properties of the original ABC-SMC algorithm. This is known as nonparametric Bayesian indirect likelihood with summary statistics, or nsBIL \citep{Drovandi2014}. \citet{Boland2017} introduced several variants of nsBIL for ABC-MCMC and proved some theoretical bounds on the approximation error.

\begin{algorithm}
\begin{algorithmic}[1]
\State Generate $\mathbf{w}_x|\beta_x$ for sample points $\beta_x$, where $x = 1,\dots,X$ \label{alg:genSw}
\State Fit the parametric functions $\hat\phi_{\sigma^2}(\beta)$ \& $\hat\phi_\mu(\beta)$ \label{alg:fitPhi}
\ForAll{iterations $t = 1, \dots, T$}
\State Update the labels $z_i \sim p(y_i | z_i, \mu_{z_i}, \sigma^2_{z_i}) \,p(z_i | z_{\setminus i}, \beta) \; \forall i \in \{ 1, \dots, n\}$ \label{alg:chequer}
\State Calculate sufficient statistics $\mathrm{S}(\mathbf{z})$ and $\bar{y}_j, s^2_j \; \forall z_i = j, \, \forall j \in \{ 1, \dots, k\}$
\State Update the noise parameters $\mu_j, \sigma^2_j \sim p(\bar{y}_j, s^2_j | \mu_j, \sigma_j^2) \,\pi(\mu_j | \sigma^2_j) \, \pi(\sigma^2_j)$ \label{alg:noise}
\State Draw proposed parameter value $\beta' \sim q(\beta' | \beta_{t-1})$
\State Approximate the Radon--Nikod{\'y}m derivative:
  \begin{equation}
  \label{eq:mhRatio_ii}
  \rho =  \frac{q(\beta_{t-1}|\beta') \pi(\beta') \, f_A\left(\mathrm{S}(\mathbf{z}) \mid \hat\phi_\mu(\beta'), \hat\phi_{\sigma^2}(\beta')\right) }{q(\beta'|\beta_{t-1}) \pi(\beta_{t-1}) \, f_A\left(\mathrm{S}(\mathbf{z}) \mid \hat\phi_\mu(\beta_{t-1}), \hat\phi_{\sigma^2}(\beta_{t-1})\right) }   \end{equation}
\State Draw $u \sim \mathrm{Uniform}[0,1]$
\If{$u < \min( 1, \rho)$}
\State $\beta_t \gets \beta'$ {\bf else} $\beta_t \gets \beta_{t-1}$
\EndIf
\EndFor
\end{algorithmic}
\caption{Parametric Functional Approximate Bayesian (PFAB) Algorithm}
\label{alg:abc-ii}
\end{algorithm}
In this section, we describe our PFAB algorithm in detail. We introduce a parametric function that takes advantage of the properties of the Potts model that were described in Section~\ref{s:model}. We utilise this function to perform Bayesian indirect inference, as shown in Algorithm~\ref{alg:abc-ii}. We have implemented this algorithm, as well as AEA and nsBIL, using \texttt{RcppArmadillo} \citep{Eddelbuettel2013}. Our open-source \textsf{R}  package for Windows, Linux or macOS is available from the CRAN repository \citep{Moores2015}.

\begin{figure}
        \centering
        \begin{subfigure}{0.85\textwidth}
                \includegraphics[width=\textwidth]{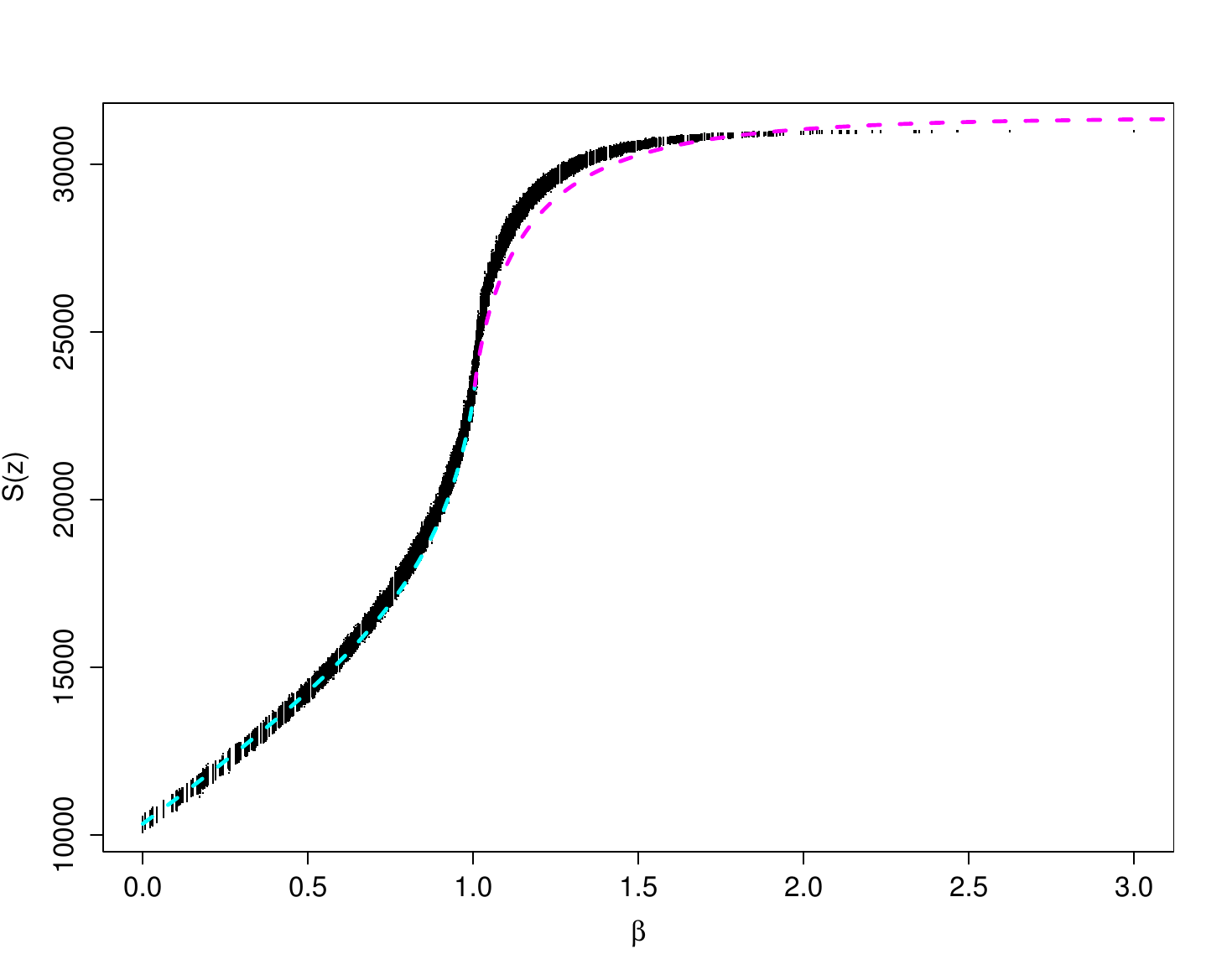}
                \caption{$\hat\phi_\mu(\beta)$}
                \label{f:paramK3n125mu}
        \end{subfigure}%
\qquad
        \begin{subfigure}{0.85\textwidth}
                \includegraphics[width=\textwidth]{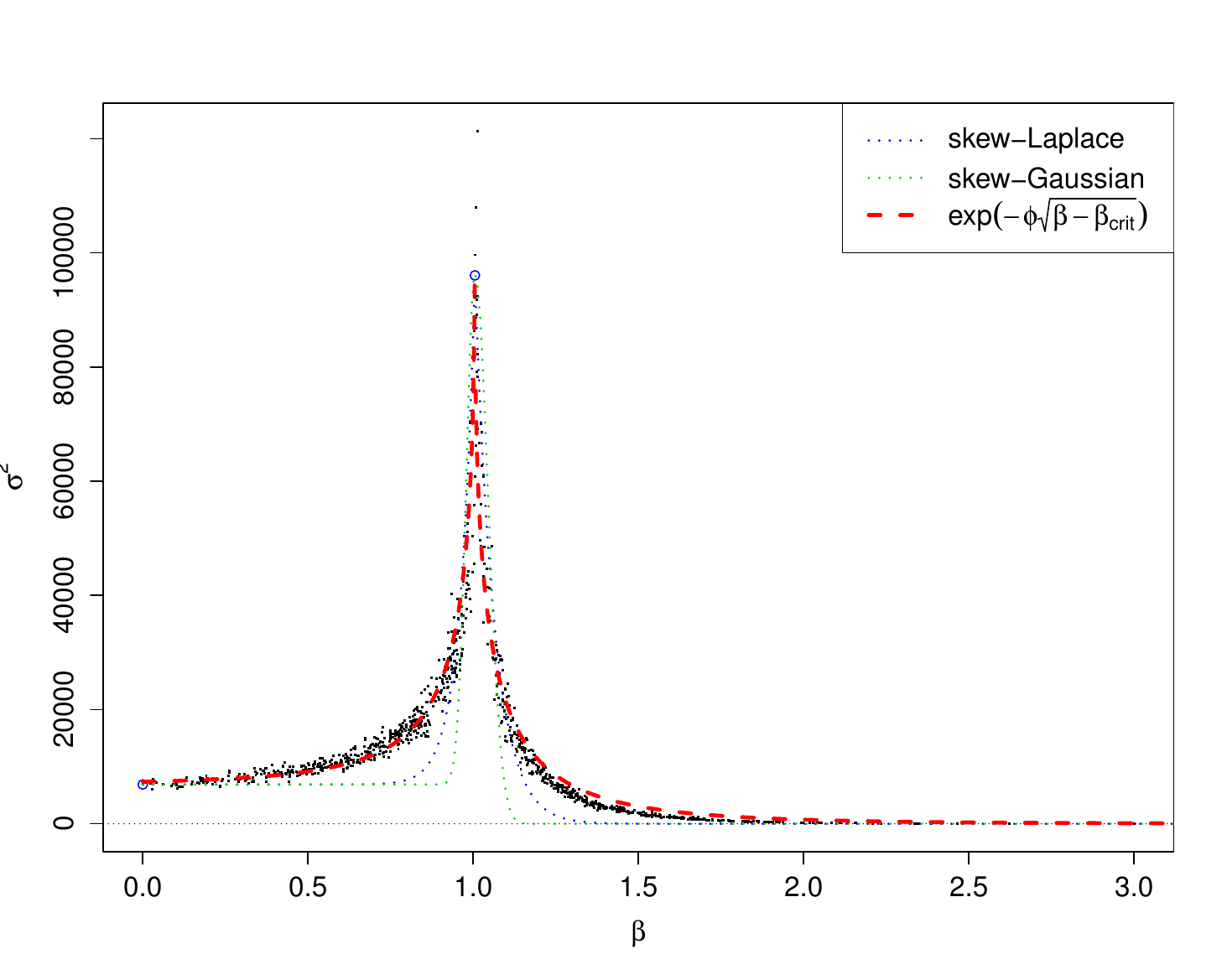}
                \caption{$\hat\phi_{\sigma^2}(\beta)$}
                \label{f:paramK3n125sd}
        \end{subfigure}
\caption{Parametric, functional approximations to $\mathrm{S}(\mathbf{w}) \;|\; \beta$ for $n=5^6, k=3$.}\label{f:pieceWise}
\end{figure}
Figures \ref{f:exact_beta} and \ref{f:bcrit} illustrate that $\mathbb{E}_{\mathbf{z} | \beta}[\mathrm{S}(\mathbf{z})]$ has the form of a sigmoid function, where the shape depends on the number of labels $k$ and edges $\#\mathcal{E}$. The variance of the distribution is equal to the gradient of this curve, according to \eqref{eq:information}. Clearly, piecewise-constant approximations such as $k$NN \citep{Sherlock2015} or random forests \citep{Pudlo2016} would be unsuitable as surrogate models. Even the piecewise-linear model of \citet{Moores2014} can create issues for MCMC algorithms, due to sharp changes in the approximate variance. At the other end of the scale, GP approximations \citep{Meeds2014} are too smooth to represent the discontinuity at the first-order phase transition boundary. Similarly, the generalised logistic function \citep{Richards1959} is not flexible enough to be used for this purpose.

 Figure~\ref{f:paramK3n125sd} illustrates three different parametric approximations to the variance of the 2D Potts model with $k=3$ and $n=5^6$. The skew-Gaussian approximation is given by:
  \begin{equation}
  \hat\phi_{\sigma^2}(\beta) =  \left\{
     \begin{array}{ll}
       \mathbb{V}_0 + (\mathbb{V}_{max} - \mathbb{V}_0) e^{-\vartheta_1 (\beta_{crit} - \beta)^2} & : 0 \le \beta < \beta_{crit},\\
       \mathbb{V}_{max} e^{-\vartheta_2 (\beta_{crit} - \beta)^2}  & : \beta \ge \beta_{crit}.
     \end{array}
   \right.
   \end{equation}
The tails of the Gaussian decay much faster than the true model. The skew-Laplace approximation is slightly better:
  \begin{equation}
  \hat\phi_{\sigma^2}(\beta) =  \left\{
     \begin{array}{ll}
       \mathbb{V}_0 + (\mathbb{V}_{max} - \mathbb{V}_0) e^{-\vartheta_1 (\beta_{crit} - \beta)} & : 0 \le \beta < \beta_{crit},\\
       \mathbb{V}_{max} e^{-\vartheta_2 (\beta - \beta_{crit})}  & : \beta \ge \beta_{crit}.
     \end{array}
   \right.
   \end{equation}
The approximation can be further improved by taking the square root of the distance between $\beta$ and its critical value:
  \begin{equation}\label{eq:phi_var}
  \hat\phi_{\sigma^2}(\beta) =  \left\{
     \begin{array}{ll}
       \mathbb{V}_0 + (\mathbb{V}_{max} - \mathbb{V}_0) e^{-\vartheta_1 \sqrt{\beta_{crit} - \beta}} & : 0 \le \beta < \beta_{crit},\\
       \mathbb{V}_{max} e^{-\vartheta_2 \sqrt{\beta - \beta_{crit}}}  & : \beta \ge \beta_{crit}.
     \end{array}
   \right.
   \end{equation}
   
\noindent Since the variance of the score function is equal to its derivative \eqref{eq:information}, it follows that the functional form for the expectation can be obtained as the integral curve of $\hat\phi_{\sigma^2}(\beta)$. Given the parametric function \eqref{eq:phi_var}, the corresponding approximation for the expectation can be obtained as the algebraic solution of this integral (see Appendix~\ref{s:app1}):
    \begin{equation}
  \hat\phi_\mu(\beta) = \left\{
     \begin{array}{ll}
  \mathbb{E}_0 + \beta\mathbb{V}_0 + \int_0^\beta (\mathbb{V}_{max} - \mathbb{V}_0) e^{-\vartheta_1 \sqrt{\beta_{crit} - t}}  dt & : 0 \le \beta < \beta_{crit},\\
       \mathbb{E}_{\beta_{crit}} + \int_{\beta_{crit}}^\beta \mathbb{V}_{max} e^{-\vartheta_2 \sqrt{t - \beta_{crit}}} dt & : \beta \ge \beta_{crit}.
     \end{array}
   \right.
   \end{equation}
   Figure~\ref{f:paramK3n125mu} illustrates $\hat\phi_\mu(\beta)$ for a simulated dataset. The initial conditions $\mathbb{E}_0$ and $\mathbb{V}_0$ can be calculated exactly for a rectangular lattice, according to Theorems \ref{thm:expSz} and \ref{thm:varSz}, respectively. \citet{Pickard1987} gives a formula for calculating $\mathbb{V}_{max}$, the variance at the critical point, in the asymptotic limit:
   \begin{equation}\label{eq:vmax}
   \lim_{\#\mathcal{E} \to \infty} \frac{\mathrm{S}(\mathbf{z}) - \mathbb{E}_{\mathbf{z} | \beta_{crit}}[\mathrm{S}(\mathbf{z})]}{\sqrt{\#\mathcal{E} \log\{\#\mathcal{E}\}}} \xrightarrow{d} \mathcal{N}\left(0, \frac{2}{\pi} \right).
   \end{equation}
We have found this formula useful as an upper bound, but a single parameter $\mathbb{V}_{max}$ does not produce a satisfactory fit to the simulated pseudo-data when $k > 4$. Instead, we split it into two separate paramaters, $\mathbb{V}_1 = \lim_{\beta \nearrow \beta_{crit}}\,\mathbb{V}_{\mathbf{z} | \beta}[\mathrm{S}(\mathbf{z})]$ and $\mathbb{V}_2 = \lim_{\beta \searrow \beta_{crit}}\,\mathbb{V}_{\mathbf{z} | \beta}[\mathrm{S}(\mathbf{z})]$. Due to the discontinuity at the first-order phase transition, we also treat $\mathbb{E}_{\beta_{crit}}$ as a free parameter. This leaves five parameters to estimate, $\boldsymbol\Theta = \{ \vartheta_1, \vartheta_2, \mathbb{V}_1, \mathbb{V}_2, \mathbb{E}_{\beta_{crit}} \}$, for any given values of $\#\mathcal{E}$ and $k > 4$, as shown in Table~\ref{t:results_Stan}.

At Step~\ref{alg:genSw} of Algorithm~\ref{alg:abc-ii} we select a set of sample points $\{ \beta_x \}_{x=1}^X$. We know from the Fisher information \eqref{eq:information} that we need more points close to $\beta_{crit}$, since that is where the variance is at its maximum. We selected $X=36$ points for the experiments in Section~\ref{s:results}, since we are able to generate pseudo-data independently, in parallel on a 36-core Intel Xeon server. This computation could easily be distributed across multiple servers if necessary, or run on a general-purpose graphical processing unit (GPU). However, we found that 36 points were sufficient to fit our parametric surrogate model. For each point, we ran 500 iterations of the Swendsen-Wang algorithm, discarding the first 125 as burn-in. When all of these parallel simulations are combined, we obtain a $36 \times 375$ matrix of $\mathrm{S}(\mathbf{w})$ values, conditional on $\beta_x$, $\#\mathcal{E}$, and $k$.

\begin{table}
\begin{tabular}{rrrrrrr}
\hline
$\#\mathcal{E}$ & $\vartheta_1$ & $\vartheta_2$ & $\mathbb{V}_1/\#\mathcal{E}$ & $\mathbb{V}_2/\#\mathcal{E}$ & $\mathbb{E}_{\beta_{crit}}/\#\mathcal{E}$ \\\hline
112 & [5.08; 5.23] & [3.50; 3.59] & [2.75; 2.87] & $\cdots$ & [0.5800; 0.5843] \\
1404 & [4.59; 4.70] & [4.22; 4.30] & [2.83; 2.99] & [4.55; 4.61] & [0.6281; 0.6321] \\
8064 & [4.57; 4.63] & [6.16; 6.27] & [2.92; 2.99] & [5.72; 5.73] & [0.7104; 0.7115] \\
31000 & [4.69; 4.72] & [6.46; 6.51] & [3.14; 3.18] & [5.63; 5.74] & [0.7304; 0.7324] \\
92880 & [4.71; 4.73] & [6.29; 6.33] &  [3.19; 3.21] & [4.83; 4.90] & [0.7573; 0.7584] \\
234612 & [4.69; 4.70] & [6.55; 6.59] & [3.18; 3.20] & [5.02; 5.08] & [0.7626; 0.7633] \\
1998000 & [4.72; 4.75] & [6.40; 6.43] & [3.25; 3.31] & [4.56; 4.60] & [0.7764; 0.7767] \\
\hline\end{tabular}
\caption{Posterior estimates for the parameters of the surrogate model for $k=5$ and $n=2^6, 3^6, 4^6, 5^6, 6^6, 7^6$, and $10^6$.} \label{t:results_Stan}
\end{table}

\begin{table}
\begin{tabular}{rrrrr}
\hline
$k$ & $\vartheta_1$ & $\vartheta_2$ & $\mathbb{V}_{max}/\#\mathcal{E}$ \\\hline
2 & [5.35; 5.38] & [5.24; 5.26] & [2.17; 2.18] \\
3 & [5.38; 5.39] & [5.68; 5.70] & [3.60; 3.62] \\
4 & [5.61; 5.63] & [6.10; 6.12] & [4.83; 4.86] \\
\hline\end{tabular}
\caption{Posterior estimates for the parameters of the surrogate model for $n=5^6$ and $k=2, 3$, and 4.} \label{t:results_StanK}
\end{table}

At Step~\ref{alg:fitPhi}, we use Stan \citep{Carpenter2017} to sample from the posterior distribution $\pi(\boldsymbol\Theta \mid \mathrm{S}(\mathbf{w}), \beta_x)$, treating our matrix of simulations as observed data. Following the central limit theorem established by \citet{Pickard1987}, we approximate the distribution $f(\mathrm{S}(\mathbf{w}) \mid \beta_x)$ using a truncated Gaussian:
\begin{equation}\label{eq:surrogate}
f_A\left(\mathrm{S}(\mathbf{w}) \mid \hat\phi_\mu(\beta_x), \hat\phi_{\sigma^2}(\beta_x)\right) \sim \mathcal{N}\left( \hat\phi_\mu(\beta_x), \hat\phi_{\sigma^2}(\beta_x) \right)\mathds{1}_{(0 \le \mathrm{S}(\mathbf{w}) \le \#\mathcal{E})}.
\end{equation}
Since $\mathrm{S}(\mathbf{w})$ is the count of like neighbours, this continuous model is not suitable for very small images, such as when $n=2^6$ ($\#\mathcal{E}=112$).  There are also problems for very large $\beta$ (e.g. $\beta > 1.5$), as the function approaches its horizontal asymptote. However, the fit rapidly improves as $n$ increases.

The 95\% highest posterior density (HPD) intervals for $\boldsymbol\Theta$ when $k=5$ and using 7 different values of $\#\mathcal{E}$ for each of our simulated datasets are shown in Table~\ref{t:results_Stan}. Ignoring the outlier in the first row, which is too small for asymptotic properties to apply, there are clear trends in the parameter values as the size of the lattice increases. It might be possible to interpolate between these parameter values for other $n$, or even to extrapolate to larger images. Table~\ref{t:results_StanK} shows HPD intervals for the surrogate model when $k \le 4$. There are only 3 free parameters to estimate here, since there is no first-order phase transition. 

In most cases, the number of iterations $T$ required for the primary chain will depend on the mixing time of the chequerboard Gibbs sampler that is used in Step~\ref{alg:chequer}. This depends on both the size of the state vector $\mathbf{z}$ and the value of $\beta$. The Gibbs sampler is not guaranteed to be ergodic for $\beta > \beta_{crit}$, so this is an important consideration when analysing real image data. 
%In Sect.~\ref{s:results}, we obtain a conservative estimate of the mixing time as part of the precomputation step and use this to guide our choice of $T$. 
Similarly, the heuristic number of 500 iterations advocated by \citet{Cucala2009,Everitt2012} for simulating $\mathbf{w}$ in AEA will be insufficient for large images when $\beta > \beta_{crit}$. \citet{Liang2010} showed that convergence of the auxiliary chain could be improved by initialising $\mathbf{w}$ at the current value of $\mathbf{z}$. Using this warm start, the number of required iterations depends on the autocorrelation time of the Markov chain, which can be an order of magnitude less than the duration of the transient phase or burn-in period. This can be combined with tempered transitions \citep{Neal2005} to further improve the acceptance rate. Our approach is to instead use the Swendsen-Wang algorithm to simulate $\mathbf{w}$ because this has better mixing properties for large values of $\beta$ \citep{Cooper1999}.

We use conjugate priors for the parameters of the observation equation \eqref{eq:obs}, so $\sigma^2_j$ and $\mu_j$ can be updated by Gibbs sampling from their conditional posterior distributions at Step~\ref{alg:noise}. This requires calculation of the sufficient statistics for each label: $n_j = \sum_{i=1}^n \delta(z_i, j)$; $\bar{y}_j = n_j^{-1} \sum_{\{i : z_i=j\}} y_i$; and $s^2_j = n_j^{-1} \sum_{\{i : z_i=j\}} (y_i - \bar{y}_j)^2$. The inverse temperature is sampled using an approximate Metropolis-within-Gibbs step. We use adaptive, Gaussian random walk proposals for $q(\beta' | \beta_{t-1}) \sim \mathcal{N}(\beta_{t-1}, \sigma^2_{MH})$, where the step size $\sigma^2_{MH}$ is tuned adaptively using a Robbins--Monro recursion \citep{Andrieu2008,Garthwaite2010}. The surrogate model is used in place of the intractable likelihood to approximate the Metropolis--Hastings ratio \eqref{eq:mhRatio_ii}.

\section{Experimental results}
\label{s:results}
\subsection{Simulation study} \label{s:simulation}
We generated synthetic data using the \textsf{R} package \texttt{PottsUtils} \citep{Feng2008,Feng2014} as explained in Section~\ref{s:data_sim}. We used informative priors
for the $j \in \{1, \dots, 5\}$ mixture components \eqref{eq:pr_mu} and a uniform prior on the interval $[0, 1.2 \beta_{crit}]$ for the inverse temperature. $\beta_{crit}$ was calculated using Equation~(\ref{eq:bcrit2d}) to be approximately equal to $1.174$. 
%We used Equation~(\ref{eq:bcrit3d}) to obtain an approximate value for $\beta_{crit}$ in 3D images of $0.552$. 
We chose an upper bound of $1.2 \beta_{crit} \approx 1.41$ because we found that simulated datasets with $\beta$ above that value tended to have pixels with fewer than 5 unique labels, so that the effective value of $k$ was different than what was originally specified. This was a particular problem with smaller images, where all of the pixels could be assigned the same label. It is impossible to estimate $\mu_j$ or $\sigma^2_j$ accurately if there are no pixels with the label $j$. The maximum value of $\beta$ also affects the runtime, since the mixing of the Gibbs sampler slows dramatically for $\beta > \beta_{crit}$.

The precomputation step, running Swendsen-Wang for 500 iterations with 36 different values of $\beta$, took less than 5 seconds for $n = 8^2$, $27^2$, $64^2$, or $125^2$. This increased slightly to 7.6 seconds for $n=216^2$, 19.3 seconds for $n=343^2$, and 169.8 seconds for $n=1000^2$. We used 4 parallel chains with 10,000 iterations of the no u-turn sampler (NUTS) per chain to fit the parametric surrogate model in Stan. Half of the iterations were discarded as burn-in. This took 5.5 minutes on a 2.8GHz Intel Core $i7$ laptop. The computational cost of fitting the surrogate model only depends on the number of iterations of Swendsen-Wang and NUTS, as well as the number of training points $S$, not on the sizes of the images involved. The parameters of the surrogate models are shown in tables \ref{t:results_Stan} and \ref{t:results_StanK}.

\begin{figure}
\centering
        \begin{subfigure}{0.45\textwidth}
                \includegraphics[width=\textwidth]{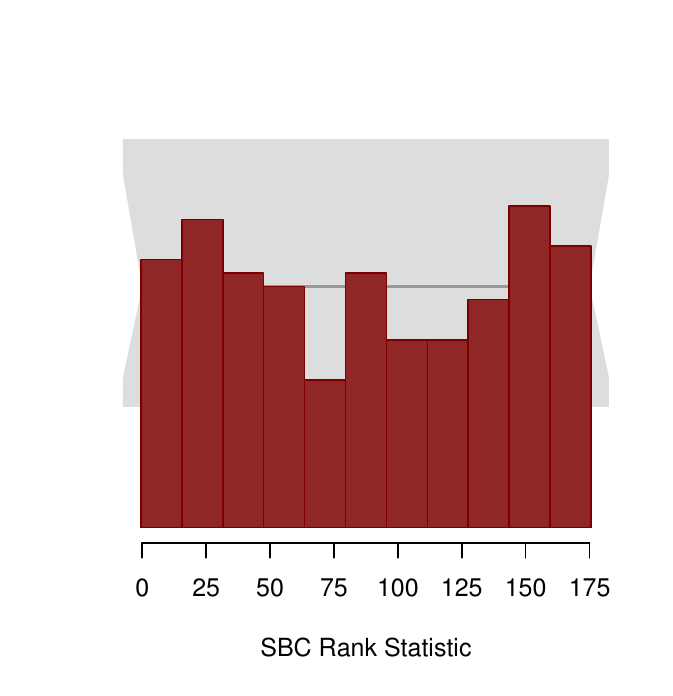}
                \caption{AEA}
        \end{subfigure}
\qquad
        \begin{subfigure}{0.45\textwidth}
                \includegraphics[width=\textwidth]{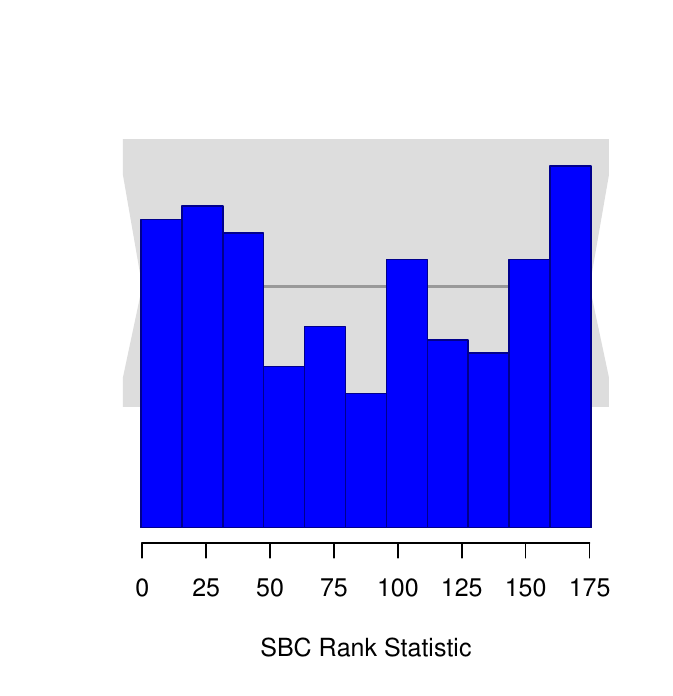}
                \caption{PFAB}
        \end{subfigure}%
\qquad
        \begin{subfigure}{0.45\textwidth}
                \includegraphics[width=\textwidth]{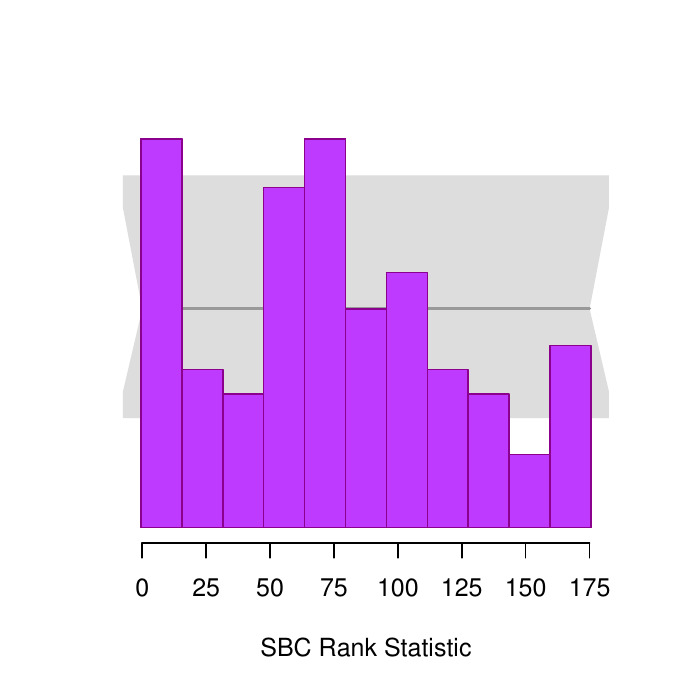}
                \caption{nsBIL}
        \end{subfigure}%
\qquad
        \begin{subfigure}{0.45\textwidth}
                \includegraphics[width=\textwidth]{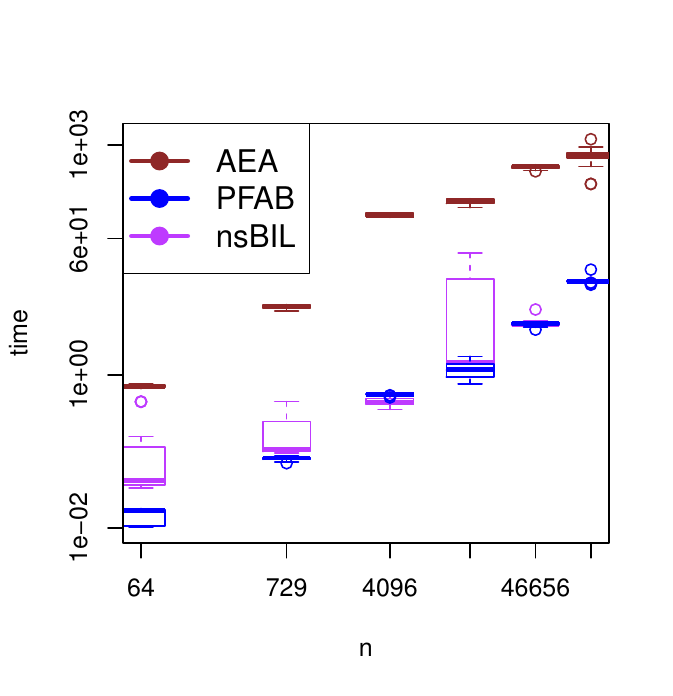}
                \caption{Elapsed runtime (minutes)} \label{f:ch6_precomp_time}
        \end{subfigure}%
\caption{Simulation-based calibration: histograms for the rank statistic of the true parameter value of $\beta$ under the posterior distribution obtained from (a) the approximate exchange algorithm; (b) our parametric functional approximate Bayesian algorithm; and (c) nonparametric Bayesian indirect likelihood with sufficient statistics. The dark grey, horizontal line shows the expected value and the light grey wedges show the 99\% confidence interval. The elapsed runtimes of the 3 algorithms are shown in (d).}
\label{f:mcmc_err}
\end{figure}

The PFAB algorithm as well as nsBIL accelerated ABC-MCMC \citep{Moores2014,Boland2017} were run for 20,000 iterations on each image, discarding the first 10,000 as burn-in. Due to the computational cost of AEA, it was run for only 5,000 iterations on the medium-sized images and 2,000 iterations on the images with 1 million pixels. The differences in elapsed runtime are illustrated by Figure~\ref{f:ch6_precomp_time}. The precomputation step for both PFAB and nsBIL-MCMC has reduced the elapsed runtime for images with $n=343^2$ pixels from 5 hours for 5,000 iterations to only 15-23 minutes for 20,000 iterations. The runtime for $n=1000^2$ pixels improved from 9-12 hours for 2,000 iterations of AEA to 2.5 hours for 20,000 iterations of PFAB or nsBIL. AEA takes less than 1 minute to run when $n=2^6$, which is less time than it takes to fit the surrogate model in Stan. Clearly, our PFAB method is not suited for small images such as these.

Histograms of the rank statistics computed using SBC \citep{Talts2018} are shown in Figure~\ref{f:mcmc_err}. The variations in these histograms of both AEA and PFAB are within the expectations of uniformity shown in the grey bars, which suggests that the fits are faithfully representing the true posterior distributions. There will naturally be some bias introduced by the surrogate model in PFAB, as well as by using Swendsen-Wang instead of perfect sampling in AEA. However, this bias is too minor to be detected by the present simulation study. On the other hand, nsBIL shows clear signs of bias towards overestimating $\beta$. This is due to a combination of the ABC tolerance $\epsilon$, as well as problems with convergence of the ABC-MCMC algorithm \citep{Lee2014}. We previously found that nsBIL performed much better when incorporated into an ABC-SMC algorithm \citep{Moores2014}, although this increases its computational cost.

\subsection{Satellite remote sensing}\label{s:satellte}
%\begin{figure}
%\centering
%        \begin{subfigure}{0.45\textwidth}
%                \includegraphics[width=\textwidth]{path_NDVI_mu.eps}
%                \caption{$\hat\phi_\mu(\beta)$}
%        \end{subfigure}
%\qquad
%        \begin{subfigure}{0.45\textwidth}
%                \includegraphics[width=\textwidth]{path_NDVI_sd.eps}
%                \caption{$\hat\phi_\sigma(\beta)$}
%        \end{subfigure}%
%\qquad
%        \begin{subfigure}{0.45\textwidth}
%                \includegraphics[width=\textwidth]{param_NDVI_mu.eps}
%                \caption{$\hat\phi_\mu(\beta)$}
%        \end{subfigure}
%\qquad
%        \begin{subfigure}{0.45\textwidth}
%                \includegraphics[width=\textwidth]{param_NDVI_sd.eps}
%                \caption{$\hat\phi_\sigma(\beta)$}
%        \end{subfigure}%
%\caption{Auxiliary models for the satellite image (2D, $n=978380, k=6$)}
%\end{figure}
For the 18 satellite images described in Section~\ref{s:data_ndvi} we used the same priors for $\mu_j$ \eqref{eq:pr_mu}, $\sigma^2_j$ \eqref{eq:pr_sd}, and $\beta$ as for the simulation study. Since these images have the same dimensions (1000 $\times$ 1000 pixels) as the largest synthetic images, we could reuse the same surrogate models for PFAB and nsBIL. The reusability of surrogate models for many datasets is a major advantage of our approach.

\begin{figure}
\centering
        \begin{subfigure}{0.9\textwidth}
                \includegraphics[width=\textwidth]{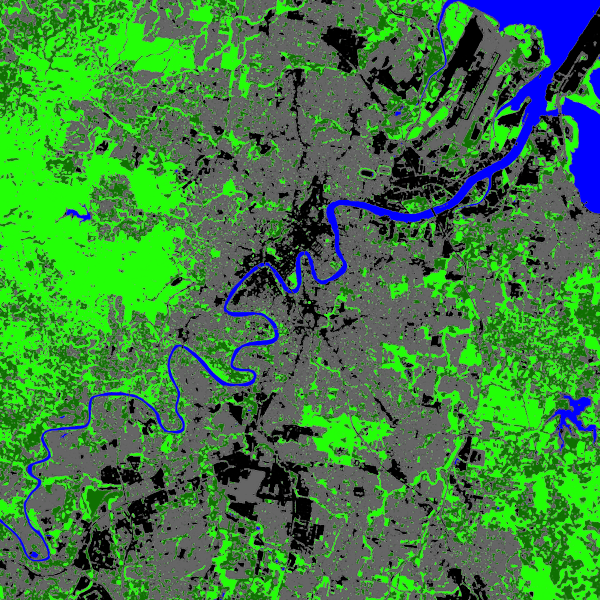}
                \caption{Image segmentation using 5 labels: water (blue); urban/industrial (black); suburbs (grey); parkland and forest (light/dark green)} \label{f:seg_ndvi}
        \end{subfigure}
\qquad
        \begin{subfigure}{0.45\textwidth}
                \includegraphics[width=\textwidth]{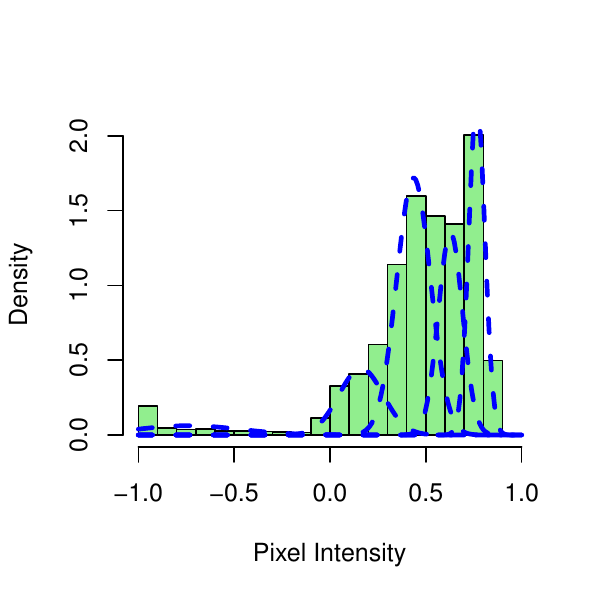}
                \caption{Mixture of Gaussians} \label{f:mix_ndvi}
        \end{subfigure}%
\qquad
        \begin{subfigure}{0.45\textwidth}
                \includegraphics[width=\textwidth]{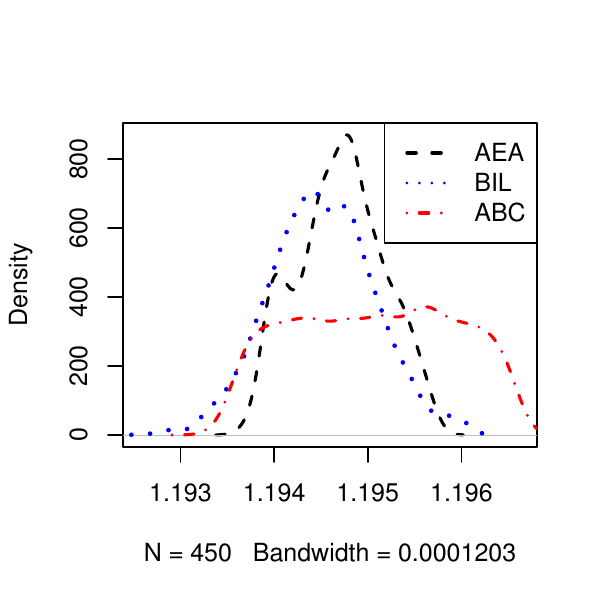}
                \caption{Posteriors for $\beta$} \label{f:post_ndvi}
        \end{subfigure}%
%\qquad
%        \begin{subfigure}{0.45\textwidth}
%                \includegraphics[width=\textwidth]{runtime_ndvi.eps}
%                \caption{Elapsed time (hours)} \label{f:runtime_ndvi}
%        \end{subfigure}%
\caption{Results for the satellite image of Brisbane, Australia.}
\label{f:results_ndvi}
\end{figure}
We were unable to evaluate accuracy for these images because the true values of the inverse temperature and the pixel labels are unknown. However, we can still compare the posterior distributions for $\beta$ obtained from the three algorithms, as well as the time taken to fit the model. These results are summarised in Figure~\ref{f:results_ndvi}. We ran PFAB and nsBIL for 20,000 iterations on each image, while AEA was run for only 2,000 iterations, with 200 iterations of Swendsen-Wang for the auxiliary variables.
% The elapsed runtimes of the algorithms are shown in Fig.~\ref{f:runtime_ndvi}. 
AEA took 19 hours per image, while PFAB and nsBIL took 1.6 hours. This resulted in an average ESS per hour of 1.88 for AEA, 598 for PFAB, and 692 for nsBIL.

An example segmentation of one of the satellite images is shown in Figure~\ref{f:seg_ndvi}. There is close correspondence between each of the 5 labels and categories of land use around Brisbane. The Brisbane River, Tingalpa Reservoir (in the south-east), and Enoggera Reservoir (in the north-west) are labelled in blue, while urbanised and industrial areas such as the central business district are labelled in black and the suburbs are labelled in grey. Large, forested areas, such as Mount Cotton and Mount Coot-tha, are labelled in green, as well as smaller features such as the City Botanic Gardens and New Farm Park. The histogram of pixel values with corresponding mixture model is shown in Figure~\ref{f:mix_ndvi}.

Figure~\ref{f:post_ndvi} illustrates the posterior density plot for $\beta$ obtained using the three algorithms. The posteriors from AEA and PFAB were generally in close agreement for the 18 images. As expected, the posterior from nsBIL is over-dispersed and uniformly-distributed over the interval $\pm \,\epsilon$. In one case, the nsBIL-MCMC algorithm failed to converge to the same value, most likely due to the issues with local modes that were observed in the simulation study.

\section{Concluding remarks}
\label{s:conclusion}
MCMC algorithms that involve simulating auxiliary variables at every iteration are too computationally expensive for applications in image analysis. Exact inference is infeasible for the scale of data that is commonly encountered in scientific studies, such as satellite remote sensing. Methods such as Russian roulette  \citep{Lyne2015}, lazy ABC \citep{Prangle2014}, or delayed acceptance \citep{Sherlock2015} can reduce the number of auxiliary iterations that need to be performed, but not enough to compensate for the two orders of magnitude difference in runtime that we observe.

To address this problem, we have introduced a parametric functional approximate Bayesian (PFAB) algorithm. This algorithm incorporates known properties of the Potts model, such as the first-order phase transition for $k > 4$, to obtain a close approximation to the true density function. We fit our surrogate model using a precomputation step, which can be run independently in parallel for selected parameter values. The same surrogate model can be reused across multiple datasets, amortising its computational cost. The resulting posterior distribution for $\beta$ has been shown to cover the true parameter value in simulation studies. When applied to real data, there is close agreement between the posterior densities obtained from PFAB and the exchange algorithm. Our approximation results in a hundred times speedup, making it feasible to use MCMC for images of realistic size, given reasonable computational power by contemporary standards. PFAB could also be used to accelerate other spatial Markov models, not only for image analysis.

The score function of the Potts model is continuous with respect to $n$ and $k$, with two notable exceptions. The first is that the phase transition only emerges for sufficiently large $n$, for example $n \ge 3^6$ when $k=5$. Secondly, the nature of the phase transition is different for $k \le 4$ and $k > 4$. This is reflected in the parameters of our surrogate model, creating the possiblility of interpolating between surrogate models for a new value of $n$. PFAB might be used in combination with trans-dimensional algorithms or other methods of inference when $k$ is unknown. Other methods for rescaling the Potts model, such as \citet{Cucala2013}, might also be applied in this context.

For the purpose of comparison, we have used random walk proposals for all three algorithms. However, the parametric surrogate model could be incorporated in an approximate Hamiltonian or Langevin algorithm~\citep{Strathmann2015,Zhang2015} to improve the efficiency of the MCMC sampler. Our surrogate model also provides an approximation to the Fisher information \eqref{eq:information}. This could be used in Bayesian optimisation to select design points for iterative refinement of the approximation, in a similar approach to \citet{Gutmann2016} or \citet{Ryan2016}. It could also be used to derive an approximate Jeffreys' prior for $\beta$.

\appendix
\section{Surrogate Model}\label{s:app1}
The parametric function $\hat\phi_{\sigma^2}(\beta)$ was introduced in Section~\ref{s:methods}, where it was remarked that the corresponding approximation for the expectation was available in closed form as the integral curve $\hat\phi_\mu(\beta) = \int_0^\beta \hat\phi_{\sigma^2}(t) \; \mathrm{d}t$. This reflects the property of the Potts model that its variance is the derivative of its expectation \eqref{eq:information}. Since $\hat\phi_{\sigma^2}(\beta)$ involves the absolute value $|\beta_{crit} - \beta|$, its integral is derived piecewise as $\int_0^{\beta_{crit}} \hat\phi_{\sigma^2}(t) dt + \int_{\beta_{crit}}^\infty \hat\phi_{\sigma^2}(t) dt$. This piecewise function also allows for the first-order phase transition for $k > 4$ states, when there is a discontinuity at the critical point. Due to this, we replace $\mathbb{V}_{max}$ with two separate parameters, $\mathbb{V}_1$ when $\beta < \beta_{crit}$, and $\mathbb{V}_2$ when $\beta \ge \beta_{crit}$:

    \begin{equation*}
  \hat\phi_\mu(\beta) = \left\{
     \begin{array}{ll}
  \mathbb{E}_0 + \beta\mathbb{V}_0 + \int_0^\beta (\mathbb{V}_1 - \mathbb{V}_0) \exp\{-\vartheta_1\sqrt{\beta_{crit} - t}\}  dt & : 0 \le \beta < \beta_{crit},\\
       \mathbb{E}_{\beta_{crit}} + \int_{\beta_{crit}}^\beta \mathbb{V}_2 \exp\{-\vartheta_2\sqrt{t - \beta_{crit}}\} dt & : \beta \ge \beta_{crit}.
     \end{array}
   \right.
   \end{equation*}
where $\mathbb{E}_{\beta_{crit}} = \int_0^{\beta_{crit}} \hat\phi_{\sigma^2}(t) \; \mathrm{d}t$ if there is no discontinuity ($k \le 4$) and is treated as a free parameter otherwise. The initial conditions $\mathbb{E}_0$ (Theorem \ref{thm:expSz}) and $\mathbb{V}_0$ (Theorem \ref{thm:varSz}) and the location of the phase transition $\beta_{crit}$ \eqref{eq:bcrit2d} are assumed known, whereas the remaining parameters $\{ \mathbb{E}_{\beta_{crit}}, \vartheta_1, \mathbb{V}_1, \vartheta_2, \mathbb{V}_2 \}$ must be estimated from simulations.

For $0 \le \beta < \beta_{crit}$:
\begin{eqnarray*}
\int_0^\beta \hat\phi_{\sigma^2}(t) dt &=& \int_0^\beta \left( \mathbb{V}_0 + (\mathbb{V}_{max} - \mathbb{V}_0) \exp\left\{-\vartheta_1\sqrt{\beta_{crit} - t}\right\} \right) dt \\
&=& \mathbb{E}_0 + \beta\mathbb{V}_0 + (\mathbb{V}_{max} - \mathbb{V}_0) \int_0^\beta \exp\left\{-\vartheta_1\sqrt{\beta_{crit} - t}\right\} dt.\\
\text{By substitution:}\\
&=& \mathbb{E}_0 + \beta\mathbb{V}_0 - (\mathbb{V}_{max} - \mathbb{V}_0) \int_{\beta_{crit}}^{\beta_{crit} - \beta} \exp\left\{-\vartheta_1 \sqrt{u}\right\} du\\
&=& \mathbb{E}_0 + \beta\mathbb{V}_0 + 2(\mathbb{V}_{max} - \mathbb{V}_0) \int_{\sqrt{\beta_{crit} - \beta}}^{\sqrt{\beta_{crit}}} e^{-\vartheta_1 s} s\; ds.\\
\text{By parts:}\\
\int_{\sqrt{\beta_{crit} - \beta}}^{\sqrt{\beta_{crit}}} e^{-\vartheta_1 s} s\; ds &=& \left[ - \frac{s}{\vartheta_1 e^{\vartheta_1 s}} \right]_{\sqrt{\beta_{crit} - \beta}}^{\sqrt{\beta_{crit}}} + \frac{1}{\vartheta_1} \int_{\sqrt{\beta_{crit} - \beta}}^{\sqrt{\beta_{crit}}} e^{-\vartheta_1 s} ds \\
&=& - \frac{1}{\vartheta_1} \left(\frac{\sqrt{\beta_{crit}}}{e^{\vartheta_1 \sqrt{\beta_{crit}}}} - \frac{\sqrt{\beta_{crit} - \beta}}{e^{\vartheta_1 \sqrt{\beta_{crit} - \beta}}} + \frac{1}{\vartheta_1 e^{\vartheta_1 \sqrt{\beta_{crit}}}}.- \frac{1}{\vartheta_1 e^{\vartheta_1 \sqrt{\beta_{crit} - \beta}}} \right) \\
&=& - \frac{1}{\vartheta_1^2} \left(\frac{\vartheta_1 \sqrt{\beta_{crit}} + 1}{e^{\vartheta_1 \sqrt{\beta_{crit}}}} - \frac{\vartheta_1 \sqrt{\beta_{crit} - \beta} + 1}{e^{\vartheta_1 \sqrt{\beta_{crit} - \beta}}} \right).\\
\int_0^\beta \hat\phi_{\sigma^2}(t) dt &=& \mathbb{E}_0 + \beta\mathbb{V}_0 - 2\frac{\mathbb{V}_{max} - \mathbb{V}_0}{\vartheta_1^2} \left(\frac{\vartheta_1 \sqrt{\beta_{crit}} + 1}{e^{\vartheta_1 \sqrt{\beta_{crit}}}} - \frac{\vartheta_1 \sqrt{\beta_{crit} - \beta} + 1}{e^{\vartheta_1 \sqrt{\beta_{crit} - \beta}}} \right). 
\end{eqnarray*}

Similarly for $\beta \ge \beta_{crit}$:
\begin{eqnarray*}
\int_0^\beta \hat\phi_{\sigma^2}(t) dt &=& \int_0^{\beta_{crit}} \hat\phi_{\sigma^2}(t) dt + \int_{\beta_{crit}}^{\beta} \hat\phi_{\sigma^2}(t) dt \\
&=& \mathbb{E}_{\beta_{crit}} +  \int_{\beta_{crit}}^{\beta} \left(  \mathbb{V}_{max} e^{-\vartheta_2 \sqrt{t - \beta_{crit}}} \right) dt. \\
\text{By substitution:}\\
&=& \mathbb{E}_{\beta_{crit}} + 2\mathbb{V}_{max} \int_{0}^{\sqrt{\beta - \beta_{crit}}} e^{-\vartheta_2 s} s\; ds.\\
\text{By parts:}\\
\int_{0}^{\sqrt{\beta - \beta_{crit}}} e^{-\vartheta_2 s} s\; ds &=& - \frac{1}{\vartheta_2^2} \left( \frac{\vartheta_2\sqrt{\beta - \beta_{crit}} + 1}{e^{\vartheta_2 \sqrt{\beta - \beta_{crit}}}} \right), \\
\int_0^\beta \hat\phi_{\sigma^2}(t) dt &=&  \mathbb{E}_{\beta_{crit}} - 2\frac{\mathbb{V}_{max}}{\vartheta_2^2} \left( \frac{\vartheta_2\sqrt{\beta - \beta_{crit}} + 1}{e^{\vartheta_2 \sqrt{\beta - \beta_{crit}}}} \right).
\end{eqnarray*}

Thus:
    \begin{equation*}
  \hat\phi_\mu(\beta) = \left\{
     \begin{array}{ll}
   \mathbb{E}_0 + \beta\mathbb{V}_0 - 2\frac{\mathbb{V}_{max} - \mathbb{V}_0}{\vartheta_1^2} \left(\frac{1 + \vartheta_1 \sqrt{\beta_{crit}}}{\exp\{\vartheta_1 \sqrt{\beta_{crit}}\}} - \frac{1 + \vartheta_1 \sqrt{\beta_{crit} - \beta}}{\exp\{\vartheta_1 \sqrt{\beta_{crit} - \beta\}}} \right)  & : 0 \le \beta < \beta_{crit},\\
       \mathbb{E}_{\beta_{crit}} - 2\frac{\mathbb{V}_{max}}{\vartheta_2^2} \left( \frac{1 + \vartheta_2\sqrt{\beta - \beta_{crit}}}{\exp\{\vartheta_2 \sqrt{\beta - \beta_{crit}}\}} \right) & : \beta \ge \beta_{crit}.
     \end{array}
   \right.
   \end{equation*}

\bibliographystyle{abbrvnat}
\bibliography{InverseTemperature}

\begin{thebibliography}{75}
\providecommand{\natexlab}[1]{#1}
\providecommand{\url}[1]{\texttt{#1}}
\expandafter\ifx\csname urlstyle\endcsname\relax
  \providecommand{\doi}[1]{doi: #1}\else
  \providecommand{\doi}{doi: \begingroup \urlstyle{rm}\Url}\fi

\bibitem[Alston et~al.(2007)Alston, Mengersen, Robert, Thompson, Littlefield,
  Perry, and Ball]{Alston2007}
C.~L. Alston, K.~L. Mengersen, C.~P. Robert, J.~M. Thompson, P.~J. Littlefield,
  D.~Perry, and A.~J. Ball.
\newblock {Bayes}ian mixture models in a longitudinal setting for analysing
  sheep {CAT} scan images.
\newblock \emph{Comput. Stat. Data Anal.}, 51:\penalty0 4282--4296, 2007.
\newblock \doi{10.1016/j.csda.2006.05.013}.

\bibitem[Andrieu and Thoms(2008)]{Andrieu2008}
C.~Andrieu and J.~Thoms.
\newblock A tutorial on adaptive mcmc.
\newblock \emph{Stat. Comput.}, 18\penalty0 (4):\penalty0 343--373, 2008.
\newblock \doi{10.1007/s11222-008-9110-y}.

\bibitem[Baxter(1973)]{Baxter1973}
R.~J. Baxter.
\newblock Potts model at the critical temperature.
\newblock \emph{J. Phys. C: Solid State Phys.}, 6\penalty0 (23):\penalty0 L445,
  1973.

\bibitem[Besag(1974)]{Besag1974}
J.~Besag.
\newblock Spatial interaction and the statistical analysis of lattice systems.
\newblock \emph{J. R. Stat. Soc. Ser. B}, 36\penalty0 (2):\penalty0 192--236,
  1974.

\bibitem[Boland et~al.(2017)Boland, Friel, and Maire]{Boland2017}
A.~Boland, N.~Friel, and F.~Maire.
\newblock Efficient mcmc for gibbs random fields using pre-computation.
\newblock \emph{arXiv preprint arXiv:1710.04093 [stat.CO]}, 2017.

\bibitem[Carpenter et~al.(2017)Carpenter, Gelman, Hoffman, Lee, Goodrich,
  Betancourt, Brubaker, Guo, Li, and Riddell]{Carpenter2017}
B.~Carpenter, A.~Gelman, M.~Hoffman, D.~Lee, B.~Goodrich, M.~Betancourt,
  M.~Brubaker, J.~Guo, P.~Li, and A.~Riddell.
\newblock Stan: A probabilistic programming language.
\newblock \emph{J. Statist. Soft.}, 76\penalty0 (1):\penalty0 1--32, 2017.
\newblock \doi{10.18637/jss.v076.i01}.

\bibitem[Christen and Fox(2005)]{Christen2005}
J.~A. Christen and C.~Fox.
\newblock Markov chain {M}onte {C}arlo using an approximation.
\newblock \emph{J. Comput. Graph. Stat.}, 14\penalty0 (4):\penalty0 795--810,
  2005.
\newblock \doi{10.1198/106186005X76983}.

\bibitem[Conrad et~al.(2016)Conrad, Marzouk, Pillai, and Smith]{Conrad2014}
P.~R. Conrad, Y.~M. Marzouk, N.~S. Pillai, and A.~Smith.
\newblock Accelerating asymptotically exact {MCMC} for computationally
  intensive models via local approximations.
\newblock \emph{J. Am. Stat. Assoc.}, 111\penalty0 (516):\penalty0 1591--1607,
  2016.

\bibitem[Cook et~al.(2006)Cook, Gelman, and Rubin]{Cook2006}
S.~R. Cook, A.~Gelman, and D.~B. Rubin.
\newblock Validation of software for {B}ayesian models using posterior
  quantiles.
\newblock \emph{J. Comput. Graph. Stat.}, 15\penalty0 (3):\penalty0 675--692,
  2006.
\newblock \doi{10.1198/106186006X136976}.

\bibitem[Cooper and Frieze(1999)]{Cooper1999}
C.~Cooper and A.~M. Frieze.
\newblock Mixing properties of the {S}wendsen-{W}ang process on classes of
  graphs.
\newblock \emph{Random Struct. Algor.}, 15\penalty0 (3--4):\penalty0 242--261,
  1999.
\newblock
  \doi{10.1002/(SICI)1098-2418(199910/12)15:3/4<242::AID-RSA4>3.0.CO;2-C}.

\bibitem[Cucala and Marin(2013)]{Cucala2013}
L.~Cucala and J.-M. Marin.
\newblock Bayesian inference on a mixture model with spatial dependence.
\newblock \emph{J. Comput. Graph. Stat.}, 22\penalty0 (3):\penalty0 584--597,
  2013.
\newblock \doi{10.1080/10618600.2013.805652}.

\bibitem[Cucala et~al.(2009)Cucala, Marin, Robert, and
  Titterington]{Cucala2009}
L.~Cucala, J.-M. Marin, C.~P. Robert, and D.~M. Titterington.
\newblock A {B}ayesian reassessment of nearest-neighbor classification.
\newblock \emph{J. Am. Stat. Assoc.}, 104\penalty0 (485):\penalty0 263--273,
  2009.
\newblock \doi{10.1198/jasa.2009.0125}.

\bibitem[Drovandi et~al.(2011)Drovandi, Pettitt, and Faddy]{Drovandi2011a}
C.~C. Drovandi, A.~N. Pettitt, and M.~J. Faddy.
\newblock Approximate {B}ayesian computation using indirect inference.
\newblock \emph{J. R. Stat. Soc. Ser. C}, 60\penalty0 (3):\penalty0 317--337,
  2011.
\newblock \doi{10.1111/j.1467-9876.2010.00747.x}.

\bibitem[Drovandi et~al.(2015)Drovandi, Pettitt, and Lee]{Drovandi2014}
C.~C. Drovandi, A.~N. Pettitt, and A.~Lee.
\newblock Bayesian indirect inference using a parametric auxiliary model.
\newblock \emph{Stat. Sci.}, 30\penalty0 (1):\penalty0 72--95, 2015.
\newblock \doi{10.1214/14-STS498}.

\bibitem[Drovandi et~al.(2018)Drovandi, Moores, and Boys]{Drovandi2015a}
C.~C. Drovandi, M.~T. Moores, and R.~J. Boys.
\newblock Accelerating pseudo-marginal {MCMC} using {G}aussian processes.
\newblock \emph{Comput. Stat. Data Anal.}, 118:\penalty0 1--17, 2018.
\newblock \doi{10.1016/j.csda.2017.09.002}.

\bibitem[Eddelbuettel and Sanderson(2014)]{Eddelbuettel2013}
D.~Eddelbuettel and C.~Sanderson.
\newblock {RcppArmadillo}: Accelerating {R} with high-performance {C++} linear
  algebra.
\newblock \emph{Comput. Stat. Data Anal.}, 71:\penalty0 1054--63, 2014.
\newblock \doi{10.1016/j.csda.2013.02.005}.

\bibitem[Everitt(2012)]{Everitt2012}
R.~G. Everitt.
\newblock Bayesian parameter estimation for latent {M}arkov random fields and
  social networks.
\newblock \emph{J. Comput. Graph. Stat.}, 21\penalty0 (4):\penalty0 940--960,
  2012.
\newblock \doi{10.1080/10618600.2012.687493}.

\bibitem[Feng(2008)]{Feng2008}
D.~Feng.
\newblock \emph{Bayesian hidden {M}arkov normal mixture models with application
  to {MRI} tissue classification}.
\newblock PhD thesis, University of Iowa, 2008.

\bibitem[Feng and Tierney(2011)]{Feng2014}
D.~Feng and L.~Tierney.
\newblock \emph{{PottsUtils}: Utility Functions of the {P}otts Models}, 2011.
\newblock URL \url{http://CRAN.R-project.org/package=PottsUtils}.
\newblock R package version 0.2-2.

\bibitem[Flood(2014)]{Flood2014}
N.~Flood.
\newblock Continuity of reflectance data between {Landsat-7 ETM+} and
  {Landsat-8 OLI}, for both top-of-atmosphere and surface reflectance: A study
  in the {A}ustralian landscape.
\newblock \emph{Remote Sens.}, 6\penalty0 (9):\penalty0 7952--7970, 2014.
\newblock \doi{10.3390/rs6097952}.

\bibitem[Friel and Rue(2007)]{Friel2007}
N.~Friel and H.~Rue.
\newblock Recursive computing and simulation-free inference for general
  factorizable models.
\newblock \emph{Biometrika}, 94\penalty0 (3):\penalty0 661--672, 2007.
\newblock \doi{10.1093/biomet/asm052}.

\bibitem[Garthwaite et~al.(2015)Garthwaite, Fan, and Sisson]{Garthwaite2010}
P.~H. Garthwaite, Y.~Fan, and S.~A. Sisson.
\newblock Adaptive optimal scaling of {M}etropolis-{H}astings algorithms using
  the {R}obbins-{M}onro process.
\newblock \emph{Commun. Stat. Theory Methods}, 45\penalty0 (17):\penalty0
  5098--5111, 2015.
\newblock \doi{10.1080/03610926.2014.936562}.

\bibitem[Gelman(2017)]{Gelman2017}
A.~Gelman.
\newblock Correction to cook, gelman, and rubin (2006).
\newblock \emph{J. Comput. Graph. Stat.}, 26:\penalty0 940, 2017.

\bibitem[Geman and Geman(1984)]{Geman1984}
S.~Geman and D.~Geman.
\newblock Stochastic relaxation, {Gibbs} distributions and the {Bayesian}
  restoration of images.
\newblock \emph{IEEE Trans. PAMI}, 6:\penalty0 721--41, 1984.

\bibitem[Geweke(2004)]{Geweke2004}
J.~Geweke.
\newblock Getting it right: Joint distribution tests of posterior simulators.
\newblock \emph{J. Am. Stat. Assoc.}, 99:\penalty0 799--804, 2004.

\bibitem[Green and Richardson(2002)]{Green2002}
P.~J. Green and S.~Richardson.
\newblock Hidden {Markov} models and disease mapping.
\newblock \emph{J. Am. Stat. Assoc.}, 97:\penalty0 1055--1070, 2002.

\bibitem[Grelaud et~al.(2009)Grelaud, Robert, Marin, Rodolphe, and
  Taly]{Grelaud2009}
A.~Grelaud, C.~P. Robert, J.-M. Marin, F.~Rodolphe, and J.-F. Taly.
\newblock {ABC} likelihood-free methods for model choice in {G}ibbs random
  fields.
\newblock \emph{Bayesian Analysis}, 4\penalty0 (2):\penalty0 317--336, 2009.
\newblock \doi{10.1214/09-BA412}.

\bibitem[Gutmann and Corander(2016)]{Gutmann2016}
M.~U. Gutmann and J.~Corander.
\newblock Bayesian optimization for likelihood-free inference of
  simulator-based statistical models.
\newblock \emph{J. Mach. Learn. Res.}, 17\penalty0 (1):\penalty0 4256--4302,
  2016.

\bibitem[Henderson et~al.(2011)Henderson, Storeygard, and Weil]{Henderson2011}
V.~Henderson, A.~Storeygard, and D.~N. Weil.
\newblock A bright idea for measuring economic growth.
\newblock \emph{Am. Econ. Rev.}, 101\penalty0 (3):\penalty0 194--199, 2011.
\newblock \doi{10.1257/aer.101.3.194}.

\bibitem[Huang(2010)]{Huang2010}
K.~Huang.
\newblock \emph{Introduction to statistical physics}.
\newblock Chapman \& Hall/CRC Press, Boca Raton, 2${}^\mathrm{nd}$ edition,
  2010.

\bibitem[Huber(2003)]{Huber2003}
M.~L. Huber.
\newblock A bounding chain for {S}wendsen-{W}ang.
\newblock \emph{Random Struct. Algor.}, 22\penalty0 (1):\penalty0 43--59, 2003.
\newblock \doi{10.1002/rsa.10071}.

\bibitem[Huber(2016)]{Huber2016}
M.~L. Huber.
\newblock \emph{Perfect Simulation}, volume 148 of \emph{Monographs on
  Statistics \& Applied Probability}.
\newblock Chapman \& Hall/CRC Press, Boca Raton, FL, 2016.

\bibitem[J{\"a}rvenp{\"a}{\"a} et~al.(2016)J{\"a}rvenp{\"a}{\"a}, Gutmann,
  Vehtari, and Marttinen]{Jaervenpaeae2016}
M.~J{\"a}rvenp{\"a}{\"a}, M.~Gutmann, A.~Vehtari, and P.~Marttinen.
\newblock Gaussian process modeling in approximate {B}ayesian computation to
  estimate horizontal gene transfer in bacteria.
\newblock \emph{arXiv:1610.06462 [stat.ML]}, 2016.
\newblock URL \url{https://arxiv.org/abs/1610.06462}.
\newblock arXiv preprint.

\bibitem[Lee and \L{}atuszy{\'n}ski(2014)]{Lee2014}
A.~Lee and K.~\L{}atuszy{\'n}ski.
\newblock Variance bounding and geometric ergodicity of {M}arkov chain {M}onte
  {C}arlo kernels for approximate {B}ayesian computation.
\newblock \emph{Biometrika}, 101\penalty0 (3):\penalty0 655--671, 2014.
\newblock \doi{10.1093/biomet/asu027}.

\bibitem[Li(2009)]{Li2009}
S.~Z. Li.
\newblock \emph{{Markov} Random Field Modeling in Image Analysis}.
\newblock Springer, Dordrecht, 3$^{\mathrm{rd}}$ edition, 2009.

\bibitem[Liang(2010)]{Liang2010}
F.~Liang.
\newblock A double {M}etropolis {H}astings sampler for spatial models with
  intractable normalizing constants.
\newblock \emph{J. Stat. Comput. Sim.}, 80\penalty0 (9):\penalty0 1007--1022,
  2010.
\newblock \doi{10.1080/00949650902882162}.

\bibitem[Liang et~al.(2016)Liang, Jin, Song, and Liu]{Liang2015}
F.~Liang, I.~H. Jin, Q.~Song, and J.~S. Liu.
\newblock An adaptive exchange algorithm for sampling from distributions with
  intractable normalizing constants.
\newblock \emph{J. Am. Stat. Assoc.}, 111\penalty0 (513):\penalty0 377--393,
  2016.
\newblock \doi{10.1080/01621459.2015.1009072}.

\bibitem[Lyne et~al.(2015)Lyne, Girolami, Atchad{\'e}, Strathmann, and
  Simpson]{Lyne2015}
A.-M. Lyne, M.~Girolami, Y.~Atchad{\'e}, H.~Strathmann, and D.~Simpson.
\newblock On {R}ussian roulette estimates for {B}ayesian inference with
  doubly-intractable likelihoods.
\newblock \emph{Statist. Sci.}, 30\penalty0 (4):\penalty0 443--467, 2015.
\newblock \doi{10.1214/15-STS523}.

\bibitem[McClain(2009)]{McClain2009}
C.~R. McClain.
\newblock A decade of satellite ocean color observations.
\newblock \emph{Ann. Rev. Mar. Sci.}, 1:\penalty0 19--42, 2009.
\newblock \doi{10.1146/annurev.marine.010908.163650}.

\bibitem[McGrory et~al.(2009)McGrory, Titterington, Reeves, and
  Pettitt]{McGrory2009}
C.~A. McGrory, D.~Titterington, R.~Reeves, and A.~N. Pettitt.
\newblock Variational {B}ayes for estimating the parameters of a hidden {P}otts
  model.
\newblock \emph{Stat. Comput.}, 19\penalty0 (3):\penalty0 329--340, 2009.
\newblock \doi{10.1007/s11222-008-9095-6}.

\bibitem[McGrory et~al.(2012)McGrory, Pettitt, Reeves, Griffin, and
  Dwyer]{McGrory2012}
C.~A. McGrory, A.~N. Pettitt, R.~Reeves, M.~Griffin, and M.~Dwyer.
\newblock Variational {B}ayes and the reduced dependence approximation for the
  autologistic model on an irregular grid with applications.
\newblock \emph{J. Comput. Graph. Stat.}, 21\penalty0 (3):\penalty0 781--796,
  2012.
\newblock \doi{10.1080/10618600.2012.632232}.

\bibitem[Meeds and Welling(2014)]{Meeds2014}
E.~Meeds and M.~Welling.
\newblock {GPS-ABC}: {G}aussian process surrogate approximate {B}ayesian
  computation.
\newblock In \emph{Proc. 30$^{th}$ Conf. UAI}, pages 593--602, Quebec City,
  Canada, 2014. AUAI Press.

\bibitem[Minvielle et~al.(2010)Minvielle, Doucet, Marrs, and
  Maskell]{Minvielle2010}
P.~Minvielle, A.~Doucet, A.~Marrs, and S.~Maskell.
\newblock A {B}ayesian approach to joint tracking and identification of
  geometric shapes in video sequences.
\newblock \emph{Image and Vision Computing}, 28\penalty0 (1):\penalty0
  111--123, 2010.
\newblock \doi{10.1016/j.imavis.2009.05.002}.

\bibitem[Mira et~al.(2001)Mira, M{\o}ller, and Roberts]{Mira2001}
A.~Mira, J.~M{\o}ller, and G.~O. Roberts.
\newblock Perfect slice samplers.
\newblock \emph{J. R. Stat. Soc. Ser. B}, 63\penalty0 (3):\penalty0 593--606,
  2001.
\newblock \doi{10.1111/1467-9868.00301}.

\bibitem[M{\o}ller et~al.(2006)M{\o}ller, Pettitt, Reeves, and
  Berthelsen]{Moeller2006}
J.~M{\o}ller, A.~N. Pettitt, R.~Reeves, and K.~K. Berthelsen.
\newblock An efficient {M}arkov chain {M}onte {C}arlo method for distributions
  with intractable normalising constants.
\newblock \emph{Biometrika}, 93\penalty0 (2):\penalty0 451--458, 2006.
\newblock \doi{10.1093/biomet/93.2.451}.

\bibitem[Monahan and Boos(1992)]{Monahan1992}
J.~F. Monahan and D.~D. Boos.
\newblock Proper likelihoods for {B}ayesian analysis.
\newblock \emph{Biometrika}, 79\penalty0 (2):\penalty0 271--278, 1992.
\newblock \doi{10.2307/2336838}.

\bibitem[Moores and Mengersen(2014)]{Moores2014a}
M.~T. Moores and K.~Mengersen.
\newblock Bayesian approaches to spatial inference: modelling and computational
  challenges and solutions.
\newblock In \emph{Proc. 33$^{rd}$ Int. Wkshp MaxEnt}, volume 1636 of \emph{AIP
  Conf. Proc.}, pages 112--117, Canberra, Australia, 2014. Amer. Inst. Physics.
\newblock \doi{10.1063/1.4903701}.

\bibitem[Moores and Mengersen(2018)]{Moores2015}
M.~T. Moores and K.~Mengersen.
\newblock {bayesImageS}: {B}ayesian methods for image segmentation using a
  {P}otts model, 2018.
\newblock URL \url{https://CRAN.R-project.org/package=bayesImageS}.
\newblock R package version 0.5-1.

\bibitem[Moores et~al.(2015)Moores, Drovandi, Mengersen, and
  Robert]{Moores2014}
M.~T. Moores, C.~C. Drovandi, K.~Mengersen, and C.~P. Robert.
\newblock Pre-processing for approximate {B}ayesian computation in image
  analysis.
\newblock \emph{Stat. Comput.}, 25\penalty0 (1):\penalty0 23--33, 2015.
\newblock \doi{10.1007/s11222-014-9525-6}.

\bibitem[Murray et~al.(2006)Murray, Ghahramani, and MacKay]{Murray2006}
I.~Murray, Z.~Ghahramani, and D.~J.~C. MacKay.
\newblock {MCMC} for doubly-intractable distributions.
\newblock In \emph{Proc. $22^{nd}$ Conf. UAI}, pages 359--366, Arlington, VA,
  2006. AUAI Press.

\bibitem[NASA(2011)]{NASA2011}
NASA.
\newblock Landsat 7 science data users handbook.
\newblock Technical report, National Aeronautics and Space Administration,
  Greenbelt, MD, 2011.
\newblock URL \url{http://landsathandbook.gsfc.nasa.gov/}.

\bibitem[Neal(2005)]{Neal2005}
R.~M. Neal.
\newblock Taking bigger {M}etropolis steps by dragging fast variables.
\newblock \emph{arXiv:math/0502099 [math.ST]}, 2005.
\newblock URL \url{https://arxiv.org/abs/math/0502099}.
\newblock arXiv preprint.

\bibitem[Pickard(1987)]{Pickard1987}
D.~K. Pickard.
\newblock Inference for discrete {M}arkov fields: The simplest nontrivial case.
\newblock \emph{J. Am. Stat. Assoc.}, 82\penalty0 (397):\penalty0 90--96, 1987.
\newblock \doi{10.1080/01621459.1987.10478394}.

\bibitem[Potts(1952)]{Potts1952}
R.~B. Potts.
\newblock Some generalized order-disorder transformations.
\newblock \emph{Proc. Camb. Philos. Soc.}, 48:\penalty0 106--9, 1952.
\newblock \doi{10.1017/S0305004100027419}.

\bibitem[Prangle(2016)]{Prangle2014}
D.~Prangle.
\newblock Lazy {ABC}.
\newblock \emph{Stat. Comput.}, 26\penalty0 (1):\penalty0 171--185, 2016.
\newblock \doi{10.1007/s11222-014-9544-3}.

\bibitem[Prangle et~al.(2014)Prangle, Blum, Popovic, and Sisson]{Prangle2014a}
D.~Prangle, M.~G.~B. Blum, G.~Popovic, and S.~A. Sisson.
\newblock Diagnostic tools for approximate {B}ayesian computation using the
  coverage property.
\newblock \emph{Aust. N. Z. J. Stat.}, 56\penalty0 (4):\penalty0 309--329,
  2014.
\newblock \doi{10.1111/anzs.12087}.

\bibitem[Propp and Wilson(1996)]{Propp1996}
J.~G. Propp and D.~B. Wilson.
\newblock Exact sampling with coupled {M}arkov chains and applications to
  statistical mechanics.
\newblock \emph{Random Struct. Algor.}, 9\penalty0 (1--2):\penalty0 223--252,
  1996.
\newblock
  \doi{10.1002/(SICI)1098-2418(199608/09)9:1/2<223::AID-RSA14>3.0.CO;2-O}.

\bibitem[Pudlo et~al.(2016)Pudlo, Marin, Estoup, Cornuet, Gautier, and
  Robert]{Pudlo2016}
P.~Pudlo, J.-M. Marin, A.~Estoup, J.-M. Cornuet, M.~Gautier, and C.~P. Robert.
\newblock Reliable {ABC} model choice via random forests.
\newblock \emph{Bioinformatics}, 32\penalty0 (6):\penalty0 859--866, 2016.
\newblock \doi{10.1093/bioinformatics/btv684}.

\bibitem[Reeves and Pettitt(2004)]{Reeves2004}
R.~Reeves and A.~N. Pettitt.
\newblock Efficient recursions for general factorisable models.
\newblock \emph{Biometrika}, 91\penalty0 (3):\penalty0 751--757, 2004.
\newblock \doi{10.1093/biomet/91.3.751}.

\bibitem[Richards(1959)]{Richards1959}
F.~J. Richards.
\newblock A flexible growth function for empirical use.
\newblock \emph{J. Exp. Bot.}, 10\penalty0 (2):\penalty0 290--301, 1959.
\newblock \doi{10.1093/jxb/10.2.290}.

\bibitem[Roy et~al.(2016)Roy, Kovalskyy, Zhang, Vermote, Yan, Kumar, and
  Egorov]{Roy2016}
D.~P. Roy, V.~Kovalskyy, H.~K. Zhang, E.~F. Vermote, L.~Yan, S.~S. Kumar, and
  A.~Egorov.
\newblock Characterization of {L}andsat 7 to {L}andsat 8 reflective wavelength
  and normalized difference vegetation index continuity.
\newblock \emph{Remote Sens. Environ.}, 185:\penalty0 57--70, 2016.
\newblock \doi{10.1016/j.rse.2015.12.024}.

\bibitem[Ryan et~al.(2016)Ryan, Drovandi, and Pettitt]{Ryan2016}
C.~M. Ryan, C.~C. Drovandi, and A.~N. Pettitt.
\newblock Optimal {B}ayesian experimental design for models with intractable
  likelihoods using indirect inference applied to biological process models.
\newblock \emph{Bayesian Anal.}, 11\penalty0 (3):\penalty0 857--883, 2016.
\newblock \doi{10.1214/15-BA977}.

\bibitem[Ryd{\'e}n and Titterington(1998)]{Ryden1998}
T.~Ryd{\'e}n and D.~M. Titterington.
\newblock Computational {B}ayesian analysis of hidden {M}arkov models.
\newblock \emph{J. Comput. Graph. Stat.}, 7\penalty0 (2):\penalty0 194--211,
  1998.
\newblock \doi{10.1080/10618600.1998.10474770}.

\bibitem[Sherlock et~al.(2017)Sherlock, Golightly, and Henderson]{Sherlock2015}
C.~Sherlock, A.~Golightly, and D.~A. Henderson.
\newblock Adaptive, delayed-acceptance {MCMC} for targets with expensive
  likelihoods.
\newblock \emph{J. Comput. Graph. Stat.}, 26\penalty0 (2):\penalty0 434--444,
  2017.

\bibitem[Simoncelli(1999)]{Simoncelli1999}
E.~P. Simoncelli.
\newblock {B}ayesian multi-scale differential optical flow.
\newblock In B.~J{\"a}hne, H.~Haussecker, and P.~Geissler, editors,
  \emph{Handbook of computer vision and applications}, volume~2, chapter~14,
  pages 397--422. Academic Press, San Diego, 1999.

\bibitem[Small(2001)]{Small2001}
C.~Small.
\newblock Estimation of urban vegetation abundance by spectral mixture
  analysis.
\newblock \emph{Int. J. Remote Sens.}, 22\penalty0 (7):\penalty0 1305--1334,
  2001.
\newblock \doi{10.1080/01431160151144369}.

\bibitem[Strathmann et~al.(2015)Strathmann, Sejdinovic, Livingstone, Szabo, and
  Gretton]{Strathmann2015}
H.~Strathmann, D.~Sejdinovic, S.~Livingstone, Z.~Szabo, and A.~Gretton.
\newblock Gradient-free {H}amiltonian {M}onte {C}arlo with efficient kernel
  exponential families.
\newblock In \emph{Adv. Neur. Inf. Proc. Sys. (NIPS)}, volume~28, pages
  955--963. 2015.

\bibitem[Swendsen and Wang(1987)]{Swendsen1987}
R.~H. Swendsen and J.-S. Wang.
\newblock Nonuniversal critical dynamics in {Monte Carlo} simulations.
\newblock \emph{Phys. Rev. Lett.}, 58:\penalty0 86--88, 1987.
\newblock \doi{10.1103/PhysRevLett.58.86}.

\bibitem[Talts et~al.(2018)Talts, Betancourt, Simpson, Vehtari, and
  Gelman]{Talts2018}
S.~Talts, M.~Betancourt, D.~Simpson, A.~Vehtari, and A.~Gelman.
\newblock Validating bayesian inference algorithms with simulation-based
  calibration.
\newblock \emph{arXiv:1804.06788 [stat.ME]}, 2018.
\newblock URL \url{https://arxiv.org/abs/1804.06788}.
\newblock arXiv preprint.

\bibitem[Tucker(1979)]{Tucker1979}
C.~J. Tucker.
\newblock Red and photographic infrared linear combinations for monitoring
  vegetation.
\newblock \emph{Remote Sens. Environ.}, 8\penalty0 (2):\penalty0 127--150,
  1979.
\newblock \doi{10.1016/0034-4257(79)90013-0}.

\bibitem[USGS(2016)]{USGS2014}
USGS.
\newblock Landsat 8 data users handbook.
\newblock Technical Report LSDS-1574, United States Geological Survey, Sioux
  Falls, SD, March 2016.
\newblock URL \url{https://landsat.usgs.gov/landsat-8-l8-data-users-handbook}.
\newblock version 2.0.

\bibitem[Vermote et~al.(2016)Vermote, Justice, Claverie, and
  Franch]{Vermote2016}
E.~Vermote, C.~Justice, M.~Claverie, and B.~Franch.
\newblock Preliminary analysis of the performance of the {Landsat 8/OLI} land
  surface reflectance product.
\newblock \emph{Remote Sens. Environ.}, 185:\penalty0 46--56, 2016.
\newblock \doi{10.1016/j.rse.2016.04.008}.

\bibitem[Wilkinson(2014)]{Wilkinson2014}
R.~D. Wilkinson.
\newblock Accelerating {ABC} methods using {G}aussian processes.
\newblock In \emph{Proc. 17$^{th}$ Int. Conf. AISTATS}, volume~33 of \emph{JMLR
  W\&CP}, pages 1015--1023, Reykjavik, Iceland, 2014. MIT Press.

\bibitem[Winkler(2003)]{Winkler2003}
G.~Winkler.
\newblock \emph{Image Analysis, Random Fields and {Markov chain {Monte} Carlo}
  Methods: A Mathematical Introduction}.
\newblock Springer-Verlag, Berlin Heidelberg, 2${}^\mathrm{nd}$ edition, 2003.

\bibitem[Zhang et~al.(2017)Zhang, Shahbaba, and Zhao]{Zhang2015}
C.~Zhang, B.~Shahbaba, and H.~Zhao.
\newblock Precomputing strategy for {H}amiltonian {M}onte {C}arlo method based
  on regularity in parameter space.
\newblock \emph{Comput. Stat.}, 32\penalty0 (1):\penalty0 253--279, 2017.
\newblock \doi{10.1007/s00180-016-0683-1}.

\end{thebibliography}

\section*{Acknowledgements}
MTM was supported by the UK EPSRC as part of the ``{\it i}-Like'' programme grant (ref: EP/K014463/1). KLM was supported by an ARC Laureate Fellowship. KLM and ANP received funding from the ARC Centre of Excellence in Mathematical and Statistical Frontiers (ACEMS). The authors thank David Firth, Christian Robert, Jean-Michel Marin, Daniel Simpson, Clair Alston-Knox, and Natalie Moores for helpful conversations during the preparation of this manuscript. We also thank the associate editor and two reviewers for their thoughtful comments and suggestions.
 Landsat imagery courtesy of NASA Goddard Space Flight Center and U.S. Geological Survey. Computational resources and services used in this work were provided by the HPC and Research Support Group, Queensland University of Technology, Brisbane, Australia.

\end{document}